\documentclass[12pt]{article}

\usepackage{times,color,bm}
\usepackage{amsmath,amssymb,amsfonts,mathrsfs,braket}
\usepackage{subfigure,epsfig,graphicx,epstopdf}
\usepackage{tikz}
\usetikzlibrary{matrix,fit}
\usepackage[colorlinks]{hyperref}
\hypersetup{linkcolor = {blue},	citecolor = {blue},	urlcolor = {blue}}
\usepackage{cite}

\newenvironment{sciabstract}{%
\begin{quote} \bf}
{\end{quote}}

\newcounter{lastnote}

\begin{document}

\title{Shining Light on Quantum Transport in Fractal Networks}
\author{Xiao-Yun Xu$^{1,2}$,  Xiao-Wei Wang$^{1,2}$, Dan-Yang Chen$^{1}$\\ C. Morais Smith$^{3, 4,5 \ast}$, Xian-Min Jin$^{1,2\ast}$ \\
\normalsize{$^1$Center for Integrated Quantum Information Technologies (IQIT), School of Physics and Astronomy}\\
\normalsize{and State Key Laboratory of Advanced Optical Communication Systems and Networks,} \\
\normalsize{Shanghai Jiao Tong University, Shanghai 200240, China}\\
\normalsize{$^2$CAS Center for Excellence and Synergetic Innovation Center in Quantum Information}\\ \normalsize{and Quantum Physics, University of Science and Technology of China, Hefei, 230026, China}\\
\normalsize{$^3$Institute for Theoretical Physics, Utrecht University, Utrecht, the Netherlands}\\
\normalsize{$^4$ Tsung-Dao Lee Institute, Shanghai 200240, China}\\
\normalsize{$^5$ Wilczek Quantum Center, School of Physics and Astronomy,} \\
\normalsize{Shanghai Jiao Tong University, Shanghai 200240, China} \\
\normalsize{$^\ast$E-mail: c.demoraissmith@uu.nl}\\
\normalsize{$^\ast$E-mail: xianmin.jin@sjtu.edu.cn}
}

\date{}

\baselineskip24pt

\maketitle

\begin{sciabstract}
Fractals are fascinating structures, not only for their aesthetic appeal, but also because they allow for the investigation of physical properties in non-integer dimensions. In these unconventional systems, a myriad of intrinsic features might come into play, such as the fractal dimension, the spectral dimension, or the fractal geometry. Despite abundant theoretical and numerical studies, experiments in fractal networks remain elusive. Here, we experimentally investigate quantum transport in fractal networks by performing continuous-time quantum walks in fractal photonic lattices with incremental propagation lengths. Photons act as the walkers and evolve in the lattices after being injected into one initial site. We unveil the transport properties through the photon evolution pattern at different propagation lengths and the analysis of the variance and the P\'olya number, which are calculated based on the probability distribution of the patterns. Contrarily to classical fractals, we observe anomalous transport governed solely by the fractal dimension. In addition,  the critical point at which there is a transition from normal to anomalous transport is highly dependent on the fractal geometry. Our experiment allows the verification of physical laws in a quantitative manner and reveals the transport dynamics with unprecedented detail, thus opening a path to the understanding of more complex quantum phenomena governed by fractality.\\
\end{sciabstract}

\noindent The word ``fractal" was coined by Mandelbrot for the description of complex structures that usually exhibit self-similarity and non-integer dimension \cite{mandelbrot1977fractals, ben2000diffusion}. Closely related to humankind, nature and science \cite{mandelbrot1983fractal, gouyet1996geometry}, fractality is not only widely embodied in common objects or scenarios, such as branching trees, fluctuactions in the stock market, or human heartbeat dynamics \cite{ivanov1999heartbeat}, but is also reflected in subtle physical properties and phenomena, such as energy spectrum \cite{bandres2016topological, FractalEnergySpectrum, HofstadterSpectrum} or the growth of copper electrodeposites \cite{brady1984growth}, thereby involving the fields of physiology \cite{goldberger2002physiology, bassingthwaighte2013physiology}, finance \cite{peters1989stock}, quantum mechanics \cite{dean2013quantumHall, wang2015quantumHall, benedetti2009quantum}, optics \cite{lasermode1999, dudley2007optics, rivera2018laser}, \textit{etc}. Inspired by these natural manifestations, the concept of fractal has triggered novel designs of materials \cite{ding2018metamaterial, fan2014stretchable}, photovoltaic, and plasmonic devices \cite{fazio2016photovoltaic, gottheim2015plasmonics, de2018plasmonic, zhu2013plasmonics}, enriching the means of material property modification or of artificial device engineering.

The role of fractality has been extensively studied in the context of {\it classical} transport or diffusion by investigating classical random walk in fractal lattices \cite{ben2000diffusion, havlin1987diffusion, blumen1984continuous}. Alexander and Orbach proposed that the spectral dimension and the Hausdorff dimension govern the classical diffusion in fractals \cite{alexander1982orbach, orbach1986dynamics}. Despite some disagreements on the nature of the diffusion laws, it is beyond doubt that fractality leads to anomalous diffusion, distinctive from regular lattices, in which the mean square displacement scales linearly \cite{ben1982diffusion,sokolov2016alternative,reis2019models}.    
However, the interplay between {\it quantum} transport and fractality is still waiting for experimental exploration, despite abundant theoretical studies at non-integer dimensions \cite{mulken2011continuous,agliari2008fractaltransport,darazs2014fractaltransport,volta2009fractaltransport} and  pioneering experiments in integer-dimensional ordered \cite{feng2019quantumtransport,tang2018quantumtransport,perets2008ballistic,peruzzo2010ballistic, tang2018ballistic}, disordered \cite{schwartz2007Anderson,Lahini2008Anderson,segev2013anderson,naether2013transition} and non-Hermitian lattices \cite{eichelkraut2013transition}. Molecular synthesis  \cite{shang2015assembling,newkome2006nanoassembly,rothemund2004assembly}, atomic manipulation techniques \cite{kempkes2019electron} and photorefractive materials \cite{jia2010fractaltransport} allow the construction of well-defined fractal structures, but these platforms are not prone to the detection of dynamical properties. Photonic lattices based on femtosecond laser direct writing techniques, on the other hand, are shown to be the ideal system to investigate quantum transport dynamics,  since it enables precise design and fabrication of three-dimensional structures \cite{osellame2012femtosecond,xu2020fswriting}.

Here, we experimentally investigate quantum transport in fractal networks via continuous-time quantum walks of single photons, which is implemented on photonic lattices with identical fractal geometry but incremental propagation lengths. With the propagation length increasing, the evolution of photons at different moments is probed, visually revealing the transport dynamics. Based on the probability distribution of the evolution patterns, we calculate the variance (the mean square displacement) and the P\'olya number \cite{darazs2010polya} to characterize the transport process. Three kinds of fractals are experimentally realized to study the interplay between quantum transport and the geometrical or fractal features of the networks.

\begin{figure*}
\centering
\includegraphics[width=1 \columnwidth]{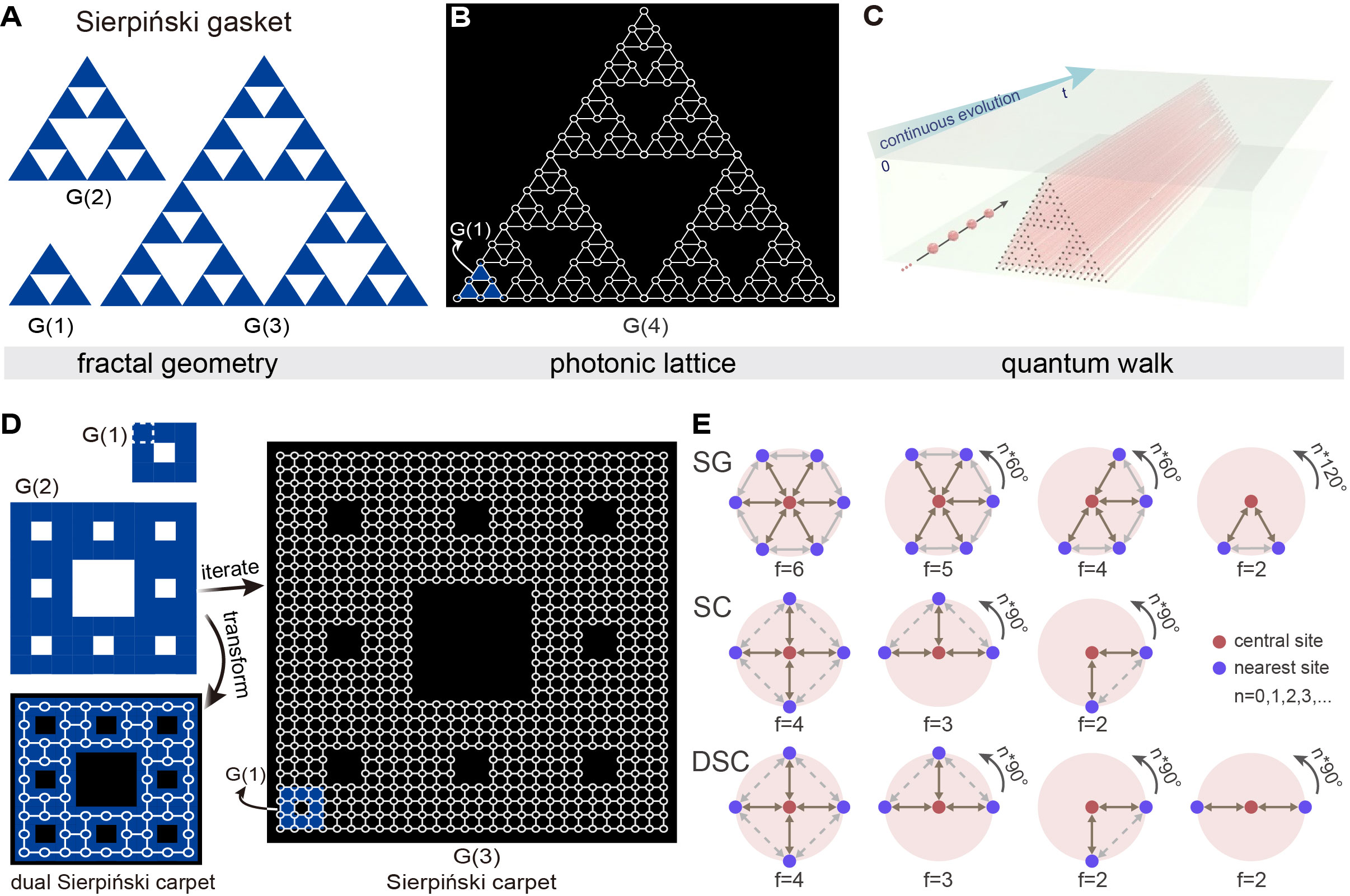}
\caption{\textbf{Geometry and connectivity analysis of the fractals, and implementation of quantum walks.} \textbf{(A)} The first-generation Sierpi\'nski gasket (SG), G(1), is constructed by removing the central part of a triangle, leaving three blue triangles. Higher generations are produced through an iterative procedure. \textbf{(B)} The photonic lattice for the fourth-generation Sierpi\'nski gasket composed of G(1) filled with blue. The white circles correspond to the corners of blue triangles and represent lattice sites. The white lines linking the circles correspond to the edges of the blue triangles and are only used for identifying the fractal geometry. \textbf{(C)} The photonic chip with embedded lattice is used to implement continuous-time quantum walks by injecting photons. The evolution time $t$ is related to the longitudinal length of the lattice. \textbf{(D)} The first-generation Sierpi\'nski carpet (SC) is constructed by removing the central part of a square, leaving eight blue squares, one of which is encircled by the white dashed line. Higher generations are formed by iterating the procedure. The white circles correspond to the corners of the blue squares, representing lattice sites. The lattice for the dual Sierpi\'nski carpet (DSC) is obtained by replacing the blue squares in the 2th-generation Sierpi\'nski carpet with lattice sites. \textbf{(E)} Connectivity analysis. Only the nearest-neighbour sites are connected (as brown or gray solid arrows denote), further neighbours are disconnected (as dashed arrows denote). The position of each site in the studied fractal lattices can be mapped to the positions of the depicted central sites or their rotations, as the angles beside the pink circular regions indicate. The connectivity of a site is equal to the number of brown solid arrows, as indicated by the value of $f$.}
\label{fig1}
\end{figure*}

Sierpi\'nski gaskets, with a Hausdorff dimension or fractal dimension of 1.58, are representative deterministic fractals \cite{darazs2014fractaltransport}. The first-generation of a Sierpi\'nski gasket  G(1) is constructed from a triangle by removing its central part, with three blue triangles left, see Fig. \hyperref[fig1]{1A}. By applying the same operation on the remaining blue triangles, we acquire the subsequent generation G(2). Similarly, higher generations are produced through an iterative procedure, revealing the self-similarity of fractals and the fact that higher generations can be decomposed into lower ones. To map the fractal geometry into photonic lattices, we define lattice sites at the corners of the blue triangles. Fig. \hyperref[fig1]{1B} exhibits the cross section of the photonic lattice for the fourth-generation Sierpi\'nski gasket, which can be regarded as a composition of identical copies of G(1) (filled with blue and denoted by the arrow). The white circles represent lattice sites and the white lines connecting the circles correspond to the edges of blue triangles. Note that the white lines are only used for identifying the fractal geometry and do not exist in the real photonic lattices. 

According to the design, we fabricate the photonic lattice depicted in Fig. \hyperref[fig1]{1C} in Corning Eagle XG glass through femtosecond laser direct writing techniques (See Methods). In light of the fact that the quantum dynamics of single photons in the lattices is equivalent to the spreading of a light beam \cite{perets2008ballistic},  horizontally polarized photons (directly coming from a coherent light beam) with a wavelength of 810 $nm$ are injected into the lattices to perform continuous-time quantum walks. Their evolution is governed by the Hamiltonian $H$ of the system,
$$|\Psi(t)\rangle=e^{-iHt}|\Psi(0)\rangle, \eqno{(1)},$$
\noindent where $|\Psi(0)\rangle$ is the initial state and $|\Psi(t)\rangle$ is the state at time $t$. Since the longitudinal propagation length $Z$ is proportional to the evolution time $t$ (i.e. $Z=vt$ where $v$ is the speed of light in the media), we can measure the evolution at different moments by preparing lattices with different lengths. 

On the assumption of tight binding, only the evanescent coupling between the nearest-neighbour sites is taken into account. The Hamiltonian is thus generally expressed as
$$H=\sum_{i=1}^{N}\beta_{i}a_{i}^{\dagger}a_{i}+\sum_{(i,j)\in E}C_{i,j}a_{i}^{\dagger}a_{j}, \eqno{(2)}$$
\noindent where $a_{i}^{\dagger}$ and $a_{i}$ are the creation and annihilation operators for photons in site $i$, $\beta_{i}$ is the propagation constant of site $i$, and $C_{i,j}$ is the coupling strength between the nearest-neighbour pair sites $i$ and $j$, as denoted by the constraint $(i,j)\in E$ where $E$ is the set composed of all the nearest-neighbour pairs of a lattice (see Methods). Therefore, the geometrical arrangement of a photonic lattice also has a notable influence on the Hamiltonian, leading to totally different quantum transport in different kinds of fractals, as we show below.

Besides the Sierpi\'nski gasket, we investigate quantum transport in other fractals with different Hausdorff dimension, namely the Sierpi\'nski carpet and its transformation, the dual Sierpi\'nski carpet, shown in Fig. \hyperref[fig1]{1D}. The first-generation Sierpi\'nski carpet, G(1), is produced by taking out the central part of a square, leaving eight blue squares, one of which is encircled by the white dashed line. The higher generations are constructed through an iteration process and the corners of the blue squares in the Sierpi\'nski carpet are defined as lattice sites. As for the dual Sierpi\'nski carpet, its photonic lattice is constructed by replacing the blue squares in the second-generation Sierpi\'nski carpet with lattice sites. Both the dual Sierpi\'nski carpet and the Sierpi\'nski carpet possess a fractal dimension of 1.89, which is different from the Sierpi\'nski gasket \cite{darazs2014fractaltransport}, but they have distinguishable geometric arrangement, such as the distribution of voids or lacunae (regions without lattice sites). The differences among these fractals allow us to study the influence on quantum transport from not only their fractal dimension, but also their geometrical features. 

We further analyze the connectivity of the fractal networks to identify their geometrical differences. Based on the tight-binding assumption, photon hopping only happens between the nearest-neighbour sites, as the solid (brown or gray) arrows in Fig. \hyperref[fig1]{1E} indicate. Only these sites are mutually connected. Otherwise, they are disconnected, as the dashed arrows denote, owing to the fact that the distance between them is larger than the nearest-neighbour length (i.e., the radius of the circular regions in pink). In contrast to regular lattices, whose sites are connected by the nearest neighbours in all directions, in the fractal lattices some nearest sites are missing, therefore resulting in a diversity of connectivity and a lack of translational invariance, which to some degree introduces the ``geometric disorder". Note that the position of each site in the studied fractal lattices can be mapped to either the position of the depicted central sites, or their rotations around the center, as the angles on the top right sides of the pink regions indicate. The connectivity of the central sites in different positions is denoted by the value of \textit{f}, which is equal to the number of brown arrows.

\begin{figure*}
\centering
\includegraphics[width=1 \columnwidth]{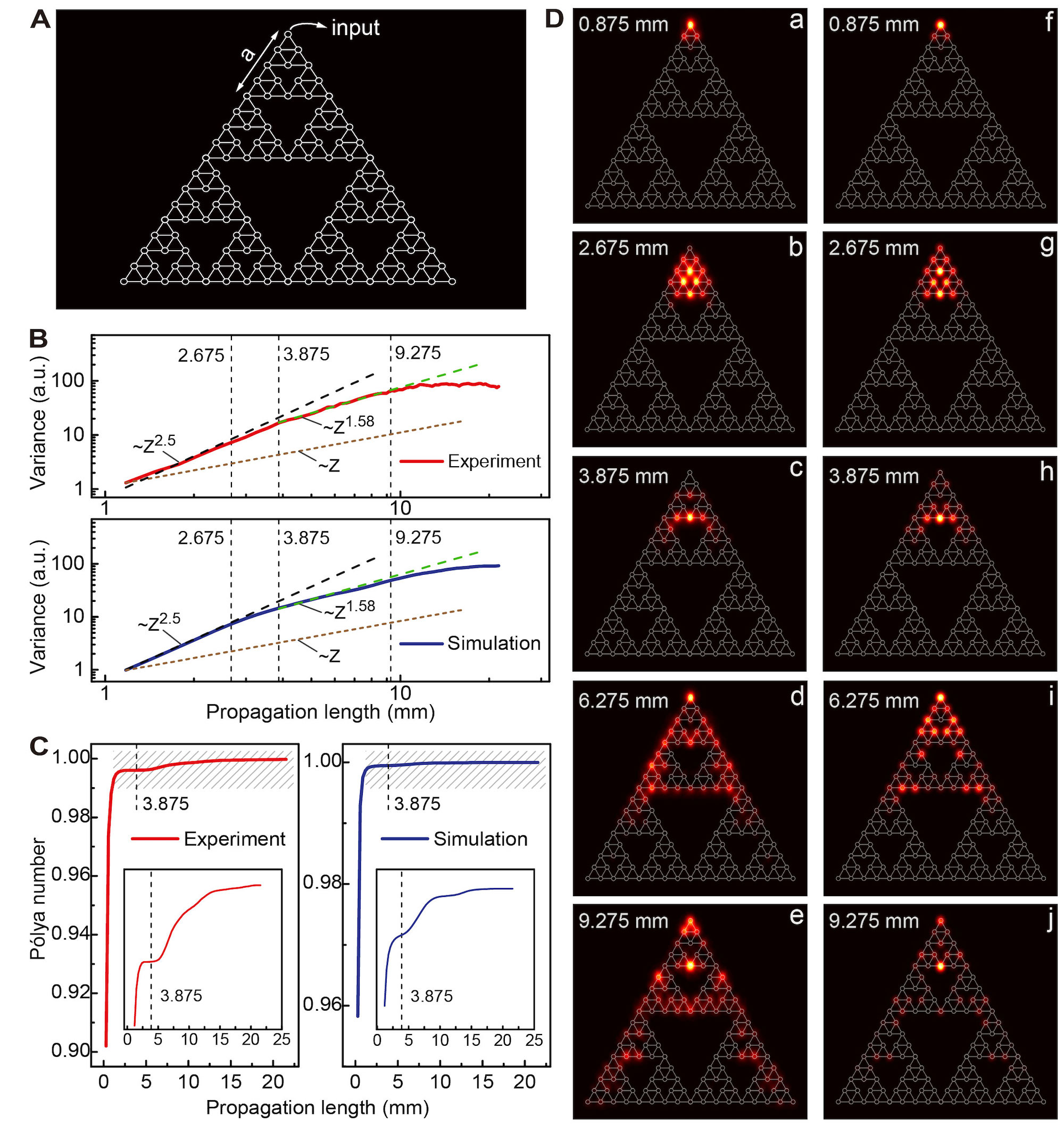}
\caption{\textbf{Quantum transport in the Sierpi\'nski gasket.} \textbf{(A)} The photonic lattice for the fourth-generation Sierpi\'nski gasket, with the apex as the input. The ``probe length" $a$ describes the transverse deviation of photons when they reach the first void. \textbf{(B)} The variance shows different scaling behavior, as the black and green dashed lines indicate. The transition points between different regimes are marked with vertical dashed lines.\textbf{(C)} The P\'olya number first grows rapidly, then goes to a plateau (before 3.875 mm). The plot highlighted by slash pattern is magnified and shown in the inset. Extra growth and plateaus appear after 3.875 mm.  \textbf{(D)} Experimental (left column) and simulated (right column) evolution patterns of photons.}
\label{fig2}
\end{figure*}

\begin{figure*}
\centering
\includegraphics[width=1\columnwidth]{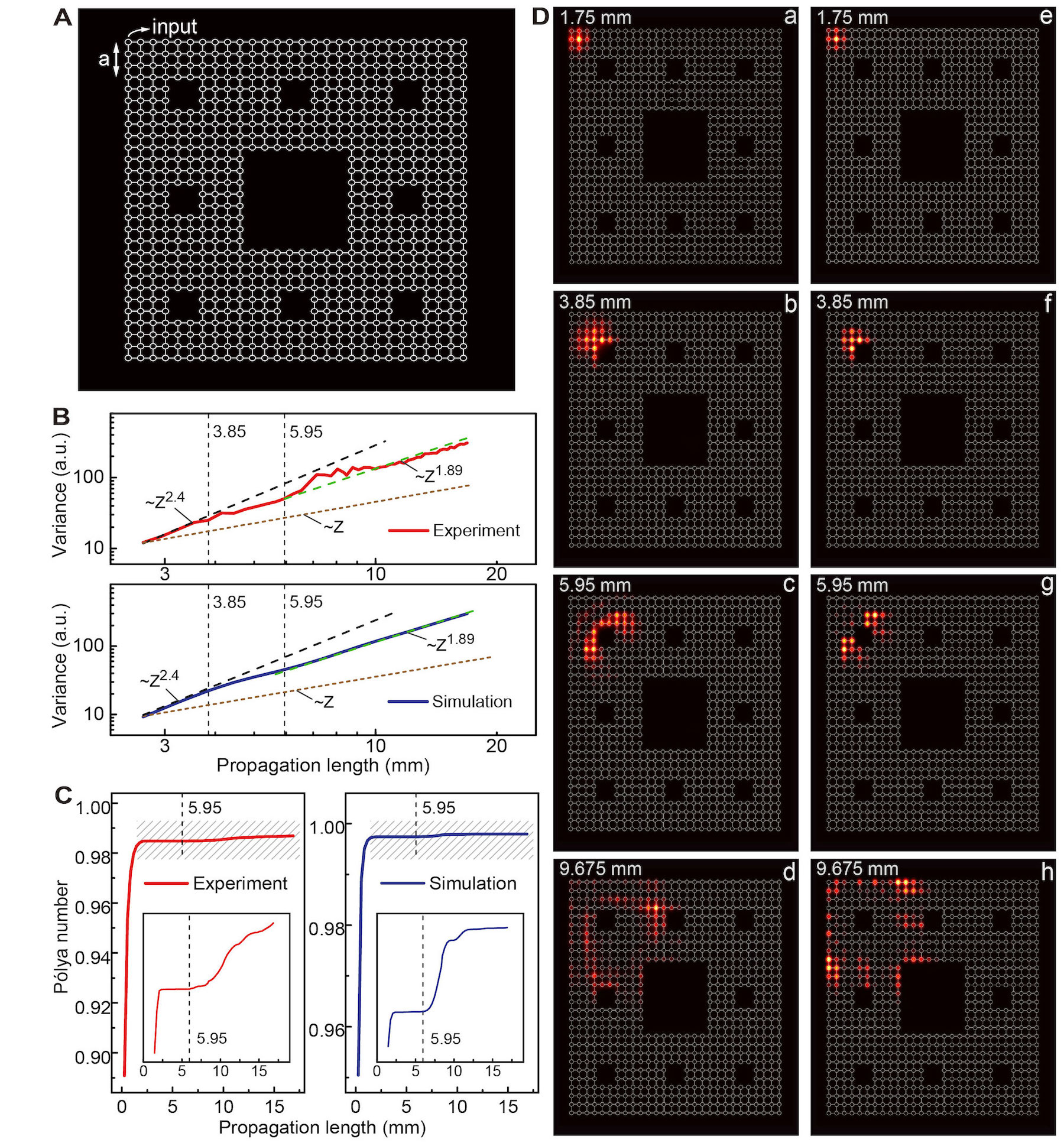}
\caption{\textbf{Quantum transport in the Sierpi\'nski carpet.} \textbf{(A)} The photonic lattice for the third-generation Sierpi\'nski carpet, with the top left corner as the input. The ``probe length" $a$ describes the transverse deviation of photons when they reach the first void.  \textbf{(B)} The variance shows different scaling behaviors, as the black and green dashed lines denote. The transition points between different regimes are marked with vertical dashed lines. \textbf{(C)} The P\'olya number first increases rapidly, then reaches a plateau (before 5.95 mm). Extra growth and plateaus are found after 5.95 mm, as shown in the inset (the magnification of the plot highlighted by slash pattern). \textbf{(D)} Experimental (left column) and simulated (right column) evolution patterns of photons.}
\label{fig3}
\end{figure*}

\begin{figure*}
\centering
\includegraphics[width=1\columnwidth]{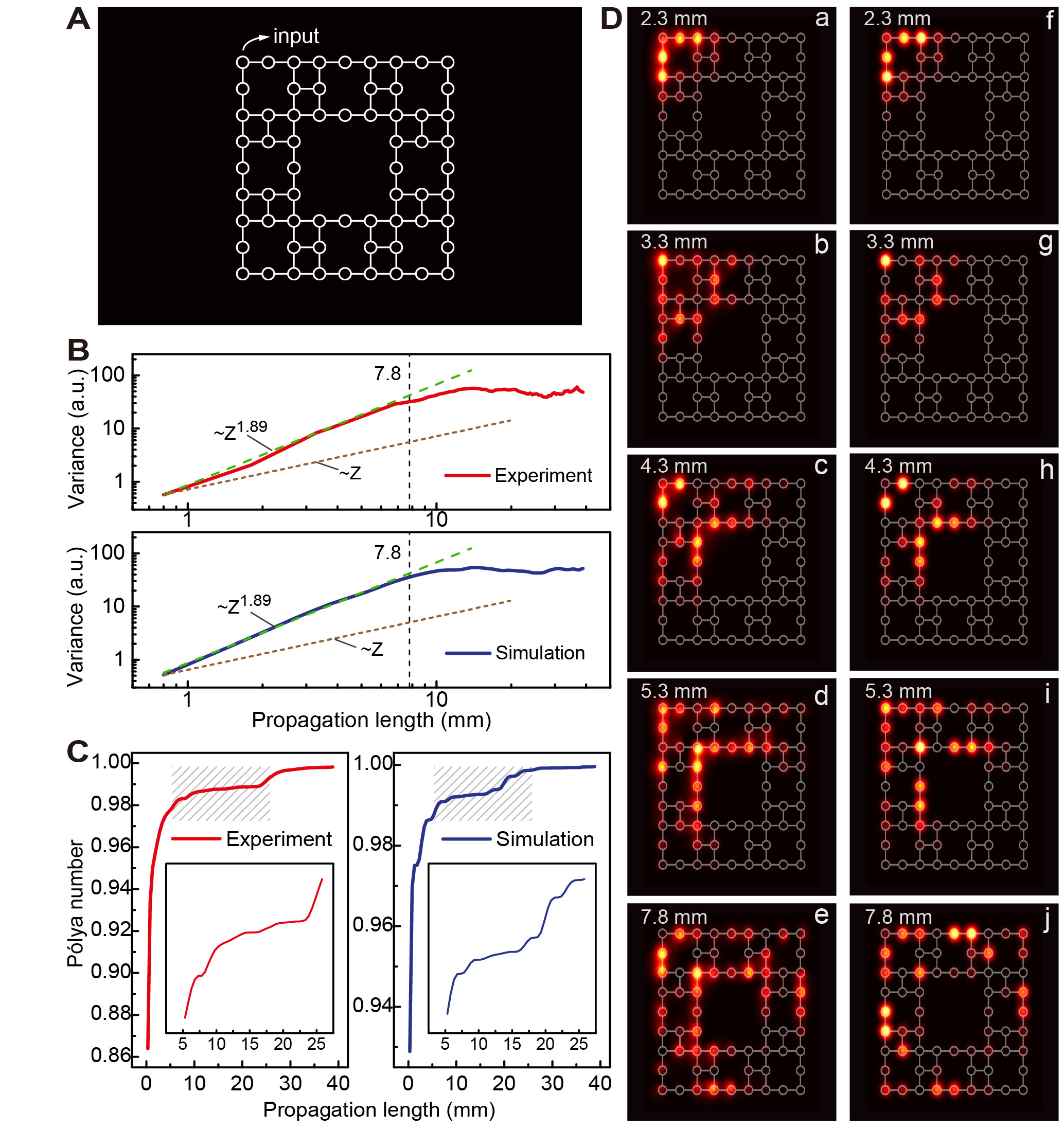}
\caption{\textbf{Quantum transport in the dual Sierpi\'nski carpet.} \textbf{(A)} The photonic lattice for the dual Sierpi\'nski carpet, a transformation of the second-generation Sierpi\'nski carpet, with the top left corner as the input. \textbf{(B)} The variance first grows as $\sim Z^{1.89}$ (as the green dashed line suggests).Then, it slowly increases and gradually saturates. The transition point is marked with vertical dashed lines. \textbf{(C)} The P\'olya number undergoes a series of growing stages and plateaus from the very beginning. The part of the plot highlighted with a slash pattern is magnified and shown in the inset. \textbf{(D)} Experimental (left column) and simulated (right column) evolution patterns of photons.}
\label{fig4}
\end{figure*}

We collect the experimental evolution patterns on the three kinds of fractal photonic lattices with single input sites as presented in Figs. \hyperref[fig2]{2A}, \hyperref[fig3]{3A} and \hyperref[fig4]{4A}, respectively. Selected experimental and simulated evolution patterns are exhibited in the left and right columns of Figs. \hyperref[fig2]{2D}, \hyperref[fig3]{3D} and \hyperref[fig4]{4D}, respectively, including those corresponding to the critical moments, which are highlighted with vertical dashed lines in Figs. \hyperref[fig2]{2B}, \hyperref[fig3]{3B} and \hyperref[fig4]{4B}. The complete evolution patterns unveiling the quantum dynamics in full are shown in the Supplementary Materials. The  variance (Figs. \hyperref[fig2]{2B}, \hyperref[fig3]{3B} and \hyperref[fig4]{4B}) and the P\'olya number (Figs. \hyperref[fig2]{2C}, \hyperref[fig3]{3C} and \hyperref[fig4]{4C}) are presented in the plots, where the propagation length $Z$ instead of time $t$ acts as the evolution coordinate. Given that $Z=vt$, it makes no difference to use either of the two coordinates.

In the case of the Sierpi\'nski gasket, the experimental and simulated variance exhibited in Fig. \hyperref[fig2]{2B} have similar trends during the entire transport process. Besides the super-diffusive transport behavior (i.e., the variance has a scaling exponent larger than 1) \cite{naether2013transition}, we find that the scaling behavior of the variance is obviously changing, yielding a curve with a variational slope in a double-logarithmic plot, which is different from the constant quadratic growth in the case of infinite regular lattices \cite{tang2018ballistic,agliari2008fractaltransport}. This anomalous phenomenon indicates a transition in the transport behavior of photons, and therefore implies a more complex regime of quantum transport in fractals than in regular media. It is also found that the fractal geometry determines the transition points of the transport. For instance, the first transition point, marked at the propagation length 2.675 mm, is exactly the critical point at which the photons reach the first void of the Sierpi\'nski gasket, as the patterns in Fig. \hyperref[fig2]{2D} reveal.

More details about the relation between the fractal geometry and the quantum transport properties are revealed in the following. For simplicity, we use ``probe length", $l_{p}$, to roughly describe the transverse deviation of photons from the input site. In general, we define that $l_{p}$ equals $a$ when the photons reach the first void of the fractal. For $0<l_{p}<a$, corresponding to $0<Z<2.675$ mm, the evolution of photons follows the normal regime describing quantum transport in finite regular lattices (see Supplementary Materials). However, the fractal geometry starts to come into play at the critical point $l_{p}=a$, or $Z=2.675$ mm. The photon transport behavior gradually deviates from the normal regime with a decrease of the scaling exponent of the variance until $l_{p}=l_{f}$ (the length at which the fractal regime starts to dominate), which is slightly larger than $a$ and corresponds to $Z=3.875$ mm. As the green dashed line indicates, the subsequent evolution of photons is governed by an anomalous or fractal regime where the variance is solely related to the fractal dimension $d_{f}$, 
$$\sigma^{2}(t) \sim t^{d_{f}},\eqno{(3)}$$ 
which is radically different from the classical transport in fractals, where the variance is governed by the ratio between the spectral dimension and the fractal dimension \cite{orbach1986dynamics,alexander1982orbach}. Moreover, the fractal regime reported here coincides with the generalized description of quantum transport in lattices with fractal energy spetrum (see Methods), despite the fact that the fractality in the two cases is manifested differently \cite{fleischmann1995quantum}. The variance keeps increasing, following the fractal regime, until photons get close to or reach the farthest site (corresponding to $Z=9.275$ mm). After that moment, the variance exhibits a saturation and small oscillations (see Methods).

The anomalies of the quantum transport are also experimentally and numerically observed in the P\'olya number displayed in Fig. \hyperref[fig2]{2C}. We magnify the plots highlighted with a slash pattern for a clearer observation and show them in the insets. The evolution of the P\'olya number undergoes a rapid growth at first and a subsequent plateau before 3.875 mm, which is the same as the P\'olya number for the regular case (see Supplementary Materials). However, we observe extra growth and plateaus, which appear alternately after 3.875 mm, as displayed in the inset. It should be noticed that the phenomenon does not appear in the regular cases and the propagation length 3.875 mm is exactly the starting point of the fractal regime, as shown in Fig. \hyperref[fig2]{2B}, which corroborates the influence of the fractal geometry on the quantum transport. 

In the case of the Sierpi\'nski carpet, as demonstrated in Figs. \hyperref[fig3]{3B} and \hyperref[fig3]{3C}, the transport properties are qualitatively similar to the ones for the Sierpi\'nski gasket, including the super-diffusive transport, the changing scaling behavior of the variance, the transition from the normal regime ($0<l_{p}<a, 0<Z<3.85$ mm ) to the fractal regime ($l_{p}>l_{f}, Z>5.95$ mm ), the decisive role of the fractal dimension in the fractal regime, and the emergence of extra growth and plateaus in the P\'olya number. The only difference, the absence of saturation and oscillation of the variance, owes to the fact that a much longer evolution time or propagation length of the photons is required to reach the farthest site of the lattice, which contains nearly 700 sites. The slight disagreement between the experiment and the simulation is due to inevitable imperfections in practical fabrications and realistic experimental environments, especially for such a large-scale lattice.

Distinctive to the above fractals, the fractal regime instead of the normal regime is probed at the very beginning of the quantum transport process in the dual Sierpi\'nski carpet. As Fig. \hyperref[fig4]{4B} presents, both the experimental and simulated variances grow as $\sim Z^{1.89}$ in the fractal regime, when $Z<7.8$ mm (before the photons hit the farthest site, as shown in Fig. \hyperref[fig4]{4D}). Subsequently, the variance increases in a very slow manner and gradually saturates. In addition, weak oscillations emerge after the saturation. The disappearance of the normal regime is attributed to the fact that the void in the dual Sierpi\'nski carpet is met from the very initial point and therefore the fractal geometry takes effect at early times. Meanwhile, as expected, a series of plateaus appear in the P\'olya number from the beginning, as demonstrated in Fig. \hyperref[fig4]{4C}. Though the dual Sierpi\'nski carpet inherits the fractal dimension of the Sierpi\'nski carpet, the transport dynamics in the two fractals has obviously distinguishable properties, reflecting that not only the fractal dimension but also other intrinsic features of the fractals play a crucial role in the quantum transport.

In summary, we have investigated quantum transport in networks of non-integer dimension, including Sierpi\'nski gaskets, Sierpi\'nski carpets and dual Sierpi\'nski carpets. With the implementation of continuous-time quantum walks in fractal photonic lattices, we revealed the transport process through either the direct observation of photon evolution patterns, or the quantitative characterization based on the variance and the P\'olya number. In contrast to the classical counterpart, different transport properties are observed, confirming the non-classical feature of the transport, which is also manifested in the return probability (see Supplementary Materials).  

Compared with infinite regular lattices, the quantum transport in fractals is anomalous. Firstly, the transport dynamics cannot be simply described by a single regime. Here, we find a crossover from the normal to the fractal regime in the cases of the Sierpi\'nski gasket and the Sierpi\'nski carpet, which is revealed by a change in the scaling behavior of the variance and by the extra growth and plateaus in the P\'olya number (after the first plateau). Secondly, the fractal features of the network play an important role in quantum transport. Indeed, the evolution of the variance in the fractal regime is completely determined by the fractal dimension. Last but not least, the  transport dynamics is also closely related to the geometrical features of the fractals. For one thing, the normal regime disappears in the case of the dual Sierpi\'nski carpet, but not in the case of the Sierpi\'nski carpet, though they have the same fractal dimension. For another, the transition points between different regimes correspond to the moments when photons reach particular positions of the fractal lattices, such as the first void and the farthest site. 

Our work sets the stage for the understanding of quantum transport in several disciplines. Given the fractal nature of organisms, e.g., trees and human bodies \cite{goldberger2002physiology}, it might shed some light on whether quantum mechanics plays any role in the transport in biological systems. The orchestrated objective reduction (Orch OR) theory proposed that consciousness originates from microtubules inside neurons, where quantum coherence emerges, and that consciousness may occur at various scales in a fractal-like brain hierarchy \cite{hameroff2014consciousness,hameroff1996consciousness}. The theory generated much controversy, with strong opponents and persuaded supporters, but the issue has not been settled. Whatever conclusion to emerge might profit from the experiments performed here. In addition, our results may also contribute to the understanding of quantum transport in disordered systems, since fractals are a good model for most disordered media \cite{havlin1987diffusion,gefen1981solvable}. Finally, our experimental demonstrations lay a strong foundation for the implementation of the quantum Grover search algorithm in fractals \cite{agliari2010algorithm}. The anomalous quantum dynamics reported here may provide new degrees of freedom for the design of quantum algorithms based on quantum walks. Therefore, by investigating quantum transport in artificially designed fractal structures, we might be making the first steps towards the long journey that will be the unification of physics, mathematics, biology, and computer science, to reach a deeper understanding of the human body and its functioning, as well as to inspire designs of more efficient quantum algorithms.

\section*{Methods}
\noindent \textbf{Preparation of fractal photonic lattices.} The fractal photonic lattices are produced by skipping over the writing of particular sites in the regular lattices corresponding to the fractals. Note that the regular lattices are constructed by filling the voids in the fractals with lattice sites. For example, the lattice for the dual Sierpi\'nski carpet or the Sierpi\'nski carpet corresponds to a square lattice with the same size. Before fabrication, based on the iterative rule for each fractal, an extra programme is used to generate a matrix whose element is either 1 or 0 to control whether a site should be written or not. Then, the matrix is imported into the fabrication programme which is originally designed for the regular lattices. Under the external control of the matrix and with the translational stage moving at a constant velocity of 10 mm/s, the lattices with desirable fractal geometry are successfully fabricated with the femtosecond laser (a repetition rate of 1 MHz, a central wavelength of 513 nm and a pulse duration of 290 fs) focused by a 50$\times$ object. Laser beam shaping through spatial light modulator and depth-dependent pulse energy compensation are applied to obtain uniform lattices. \\

\noindent \textbf{The Hamiltonian of fractal photonic lattices.} In a uniform lattice, the propagation constant for each site is equal (i.e., $\beta_{i}=\beta, i=1,2,...,N$). Since the coupling coefficient is direction-dependent, the site spacings are properly selected to ensure the same coupling coefficients for all the nearest-neighbour pairs (i.e., $C_{i,j}=C, (i,j)\in E$), as done in our previous research \cite{tang2018ballistic}. Similar to the fabrication procedure, the Hamiltonian of fractal photonic lattices is constructed from the Hamiltonian of their corresponding regular lattices. The specific procedure is described as follows. First of all, we uniquely index each site in the regular lattices. For a lattice containing $N$ sites, a $N\times N$ Hamiltonian matrix is acquired. Secondly, all the diagonal elements of the Hamiltonian are set as $\beta$. Besides, based on the geometry of the regular lattices and tight-binding assumption, the non-diagonal elements are set to be either $C$ or zero. Finally, according to the iterative rule of the fractals, the sites located in the region corresponding to the voids in the fractals are picked out and then all the coupling coefficients related to these sites are set as zero to obtain the Hamiltonian for the fractals.\\

\noindent \textbf{Simulation of photon evolution.} The simulated results of photon evolution are obtained by solving equation (1). For a lattice containing $N$ sites, the state localized at site $j$ is represented by  $|j\rangle$, and the states $\{|j\rangle, j=1,2,...,N\}$ provide an orthonormal basis set \cite{agliari2008fractaltransport}. Therefore, the system state at time $t$ can be written as  $|\Psi(t)\rangle=\sum_{j=1}^{N}a_{j}(t)|j\rangle$, where $|a_{j}(t)|^{2}=p_{j}(t)$, i.e., the probability of photons being found at site $j$ at time $t$.  In our case, only one site is initially excited (i.e.,there is only one input for photons) and therefore the initial state is denoted as $|\Psi(0)\rangle=|1\rangle$.  As a consequence,  equation (1) can be expressed as $$\sum_{j=1}^{N}a_{j}(t)|j\rangle=e^{-iHt}|1\rangle. \eqno{(4)}$$  Then we acquire that $|a_{j}(t)|^{2}=|\langle j|U(t)|1\rangle|^{2}=|U_{j,1}(t)|^{2},$ where $U(t)=e^{-iHt}$.  Based on the Hamiltonian of the system,  we calculate $U(t)$ through a Python function to obtain the probability of photons being found at each site and then display it in the form of Gaussian spots. \\

\noindent \textbf{The fractal regime of quantum transport.} As proposed by the theoretical research \cite{fleischmann1995quantum}, in the case of lattices with fractal energy spectrum, the variance is generalized as $$\sigma^{2}(t) \sim t^{2d_{f}/D}, \eqno{(5)}$$
\noindent where $D$ is the Euclidean dimension in which the lattice is embedded. For a lattice embedded in two-dimensional space, equation (5) can be written as $\sigma^{2}(t) \sim t^{d_{f}},$ which is the same as the fractal regime in our case, where the lattices show fractality in geometry.\\

\noindent \textbf{The saturation and oscillation of the variance.} The definition of variance is described by the equation
$$\sigma^2(t)=\frac{\sum_{j=1}^{N}\Delta l_{j}^{2}p_{j}(t)}{\sum_{j=1}^{N}p_{j}(t)}, \eqno{(6)}$$
\noindent where $\Delta l_{j}$ is the normalized distance between the input site and site $j$. The variance is determined by the transverse deviation and the probability distribution of photons. As photons transversely travel farther, the variance becomes larger. Meanwhile, when it is more possible to find a photon at sites far from the input, the variance tends to be larger. Therefore, as long as photons spread towards the farthest place, the variance keeps growing. The growth holds even when photons encounter the voids of the fractals, because the forward spreading is not completely stopped by the voids, though it is indeed slowed down. However, once the farthest site is reached or almost reached, the variance starts to increase very slowly and then gradually saturates. It is the finite size of the lattices that makes the saturation inevitable. The appearance of small oscillations could be a result of the complex interaction among the forward wave, the reflected wave from the farthest site and from the voids scattered across the fractals, which is still an open problem and might require further investigations.

\subsection*{Acknowledgments}
The authors thank Xuan-Lun Huang and Zhan-Ming Li for helping in the experiment, Jun Gao, Ruo-Jing Ren and Saoirse Freeney for proof reading, and Wen-Hao Zhou for assistance in formatting the figures. This research was supported by the National Key R\&D Program of China (2019YFA0308700, 2017YFA0303700), the National Natural Science Foundation of China (61734005, 11761141014, 11690033), the Science and Technology Commission of Shanghai Municipality (STCSM) (17JC1400403), the Shanghai Municipal Education Commission (SMEC) (2017-01-07-00-02- E00049). X.-M.J. acknowledges additional support from a Shanghai talent program. 

\subsection*{Author contributions} 
X.-M.J. conceived and supervised the project. X.-Y.X. performed the simulations and fabricated the photonic chips. X.-Y.X., X.-W.W., D.-Y.C. and X.-M.J. performed the experiment and analyzed the data. X.-Y.X., C.M.S. and X.-M.J. interpreted the data and wrote the paper, with input from all the other authors.

\subsection*{Competing interests}
The authors declare no competing interests.

\subsection*{Data availability}
The data that support the findings of this study are available from the corresponding authors on reasonable request.

\baselineskip21pt

\newpage


\topmargin 0.0cm  
\oddsidemargin 0.2cm
\textwidth 16cm
\textheight 22cm 
\footskip 1.0cm  

\renewcommand{\thesection}{S\Alph{section}}
\renewcommand{\thefigure}{S\arabic{figure}}
\renewcommand{\thetable}{S\Roman{table}}
\setcounter{figure}{0}
\renewcommand{\theequation}{S\arabic{equation}}

\section*{Supplementary Materials: Shining Light on Quantum Transport in Fractal Networks}

\baselineskip24pt
\subsection{Evolution patterns of photons in the fractal networks}
\begin{figure*}[htp]
\centering
\includegraphics[width=1 \columnwidth]{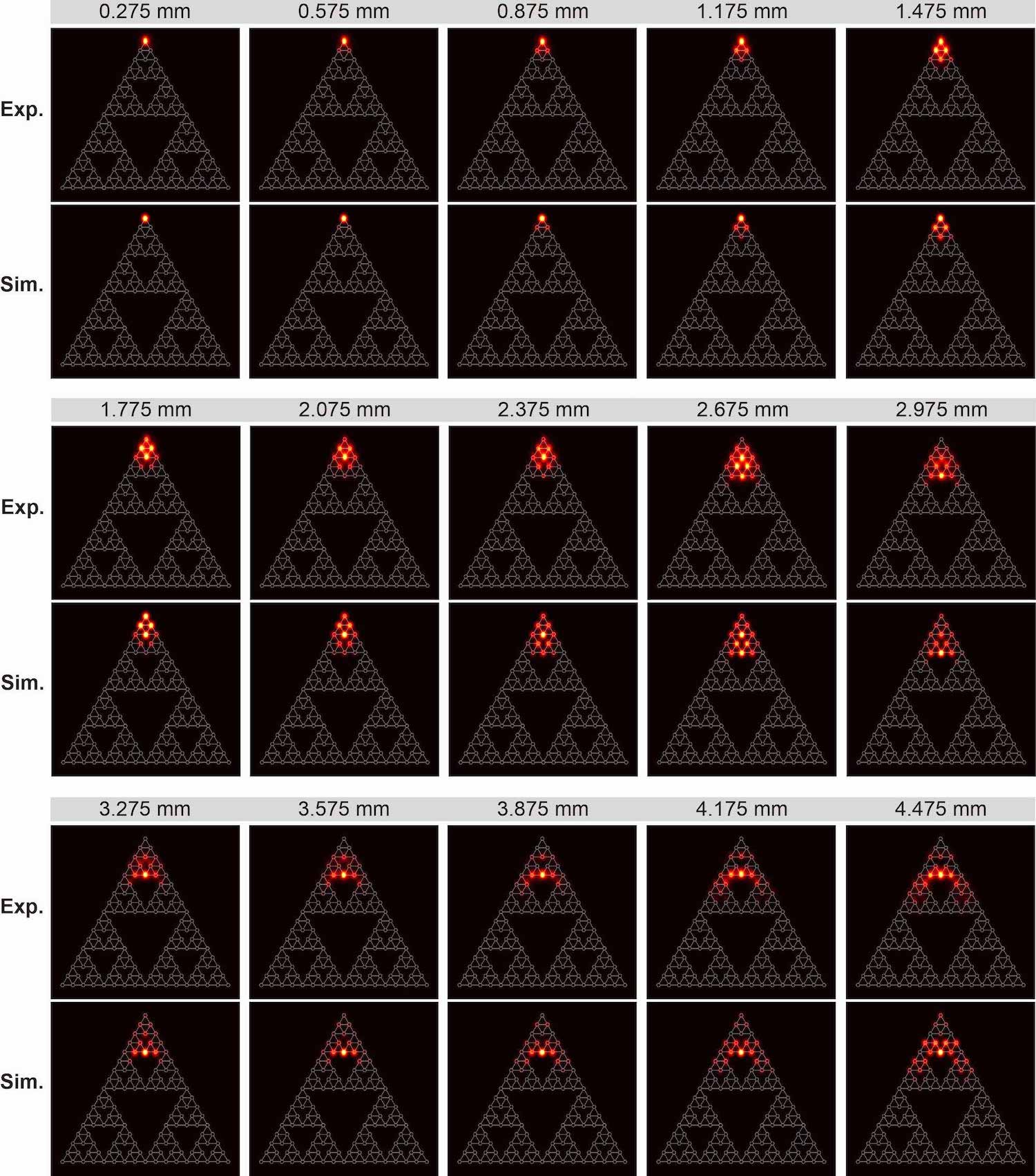}
\caption{\textbf{The evolution patterns in the Sierpi\'nski  gasket.} The propagation length starts from 0.275 mm to 4.475 mm, increasing with an increment of 0.3 mm. Experimental (Exp.) and simulated (Sim.) results are compared.}
\label{figS1}
\end{figure*}

\begin{figure*}[htp]
\centering
\includegraphics[width=1 \columnwidth]{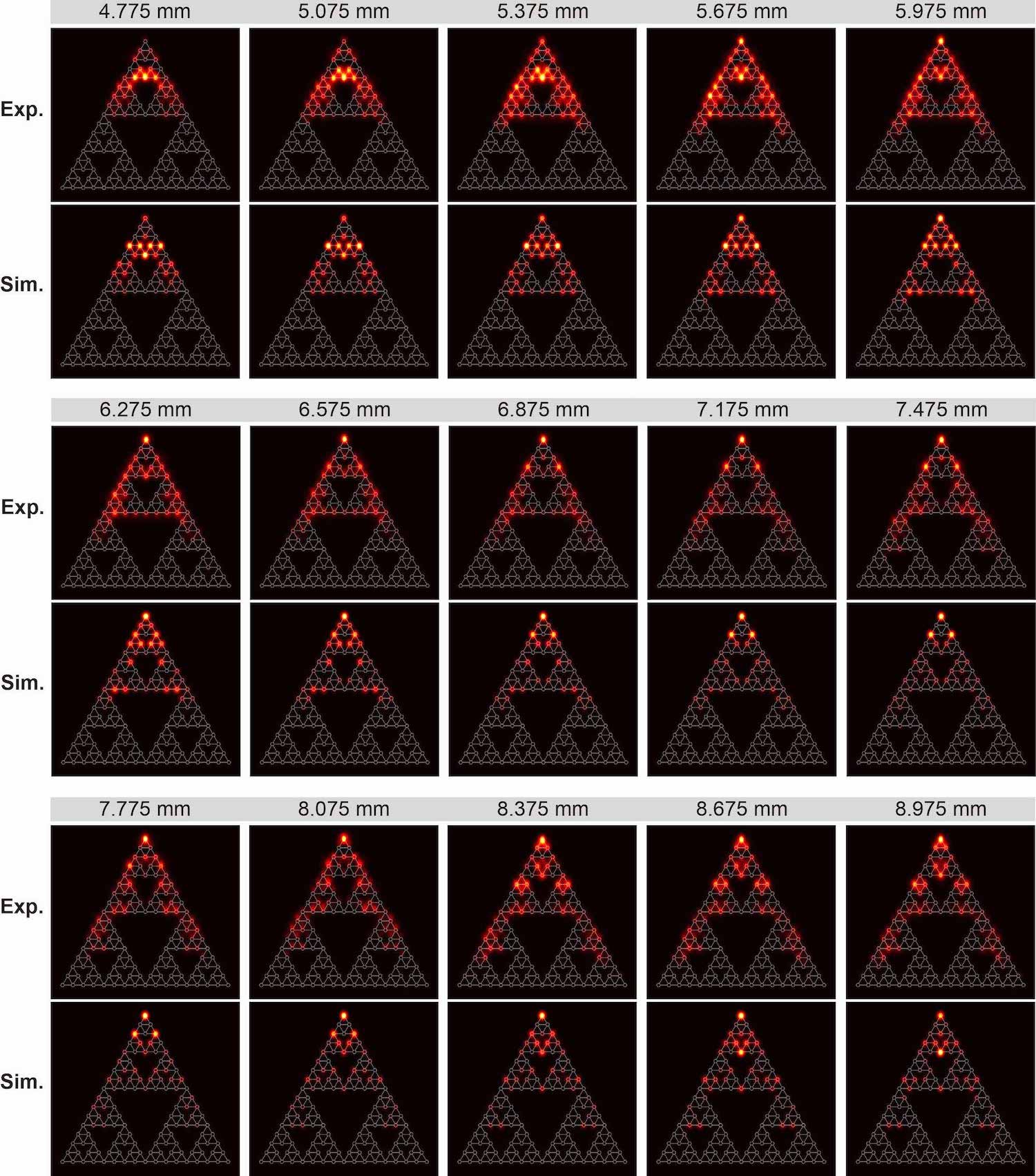}
\caption{\textbf{The evolution patterns in the Sierpi\'nski  gasket.} The propagation length starts from 4.775 mm to 8.975 mm, increasing with an increment of 0.3 mm. Experimental (Exp.) and simulated (Sim.) results are compared.}
\label{figs2}
\end{figure*}

\begin{figure*}[htp]
\centering
\includegraphics[width=1\columnwidth]{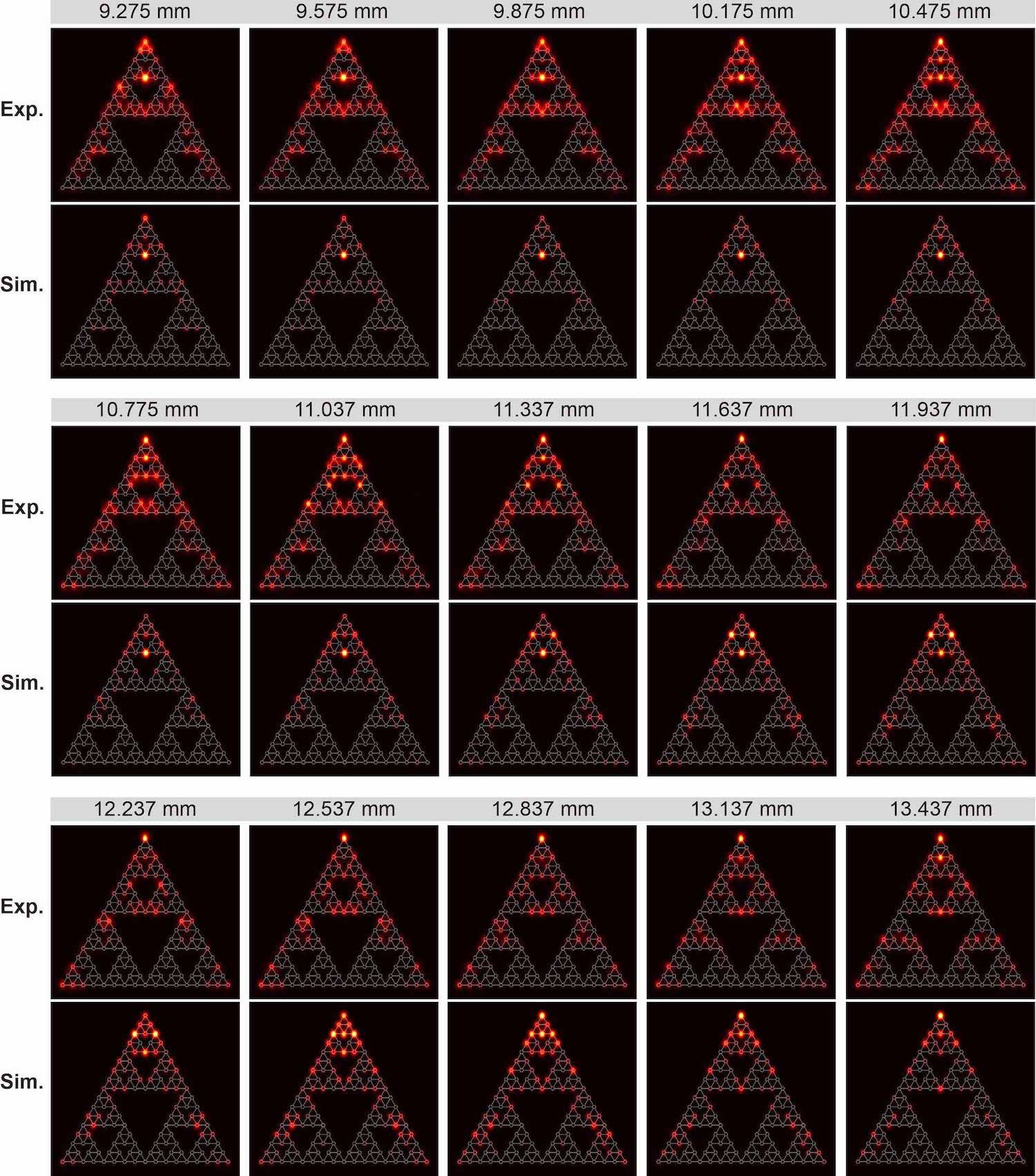}
\caption{\textbf{The evolution patterns in the Sierpi\'nski  gasket.} The propagation length starts from 9.275 mm to 13.437 mm, and increases regularly with an increment of 0.3 mm when it is smaller than 10.775 mm and larger than 11.037 mm. The  break of regular increase at 10.775 mm is attributed to the fact that the photonic chips suffer different losses in length during the mechanical grinding processes. Experimental (Exp.) and simulated (Sim.) results are compared. }
\label{figS3}
\end{figure*}

\begin{figure*}[htp]
\centering
\includegraphics[width=1\columnwidth]{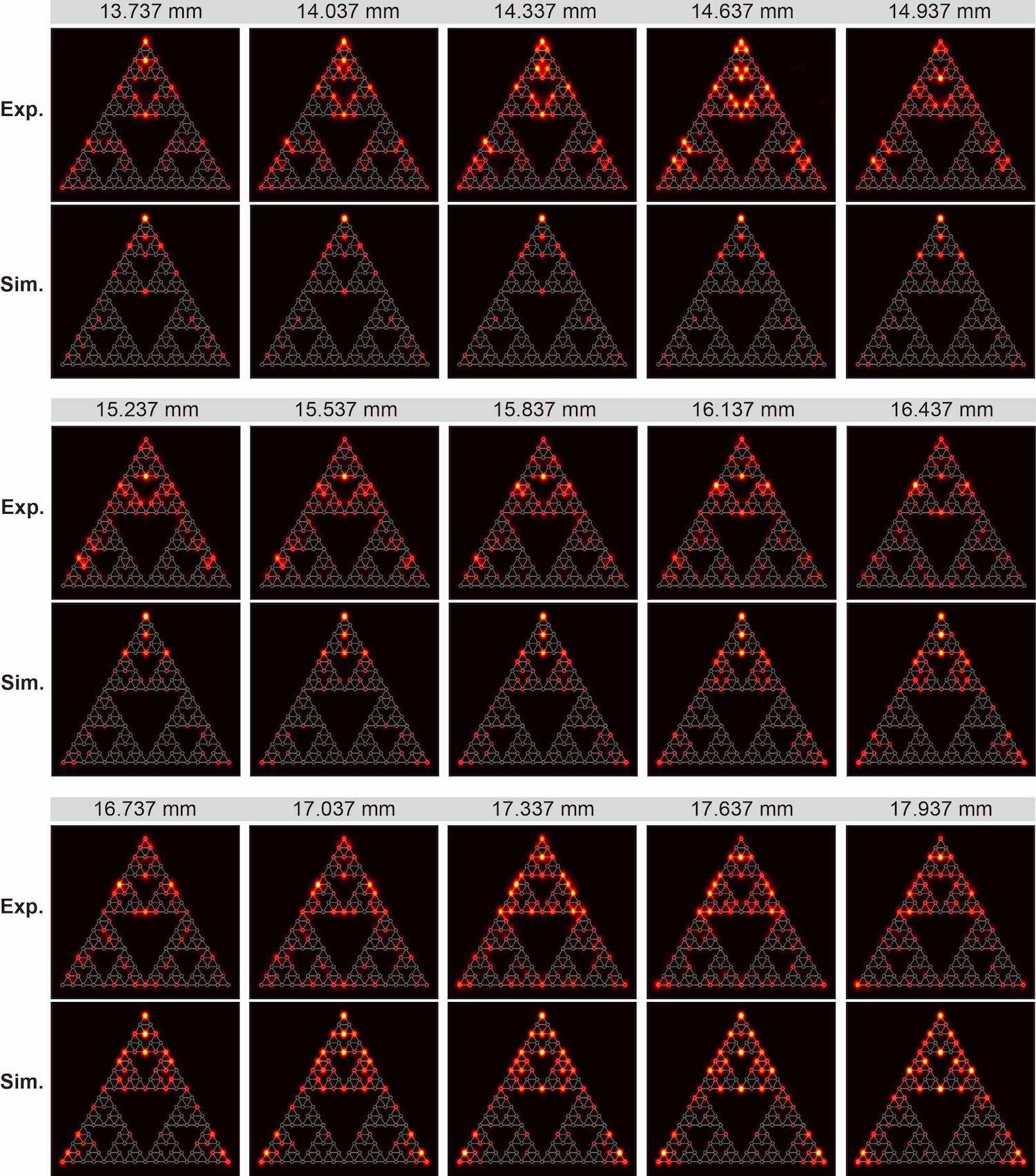}
\caption{\textbf{The evolution patterns in the Sierpi\'nski  gasket.} The propagation length starts from 13.737 mm to 17.937 mm, increasing with an increment of 0.3 mm. Experimental (Exp.) and simulated (Sim.) results are compared.}
\label{figS4}
\end{figure*}

\begin{figure*}[htp]
\centering
\includegraphics[width=1\columnwidth]{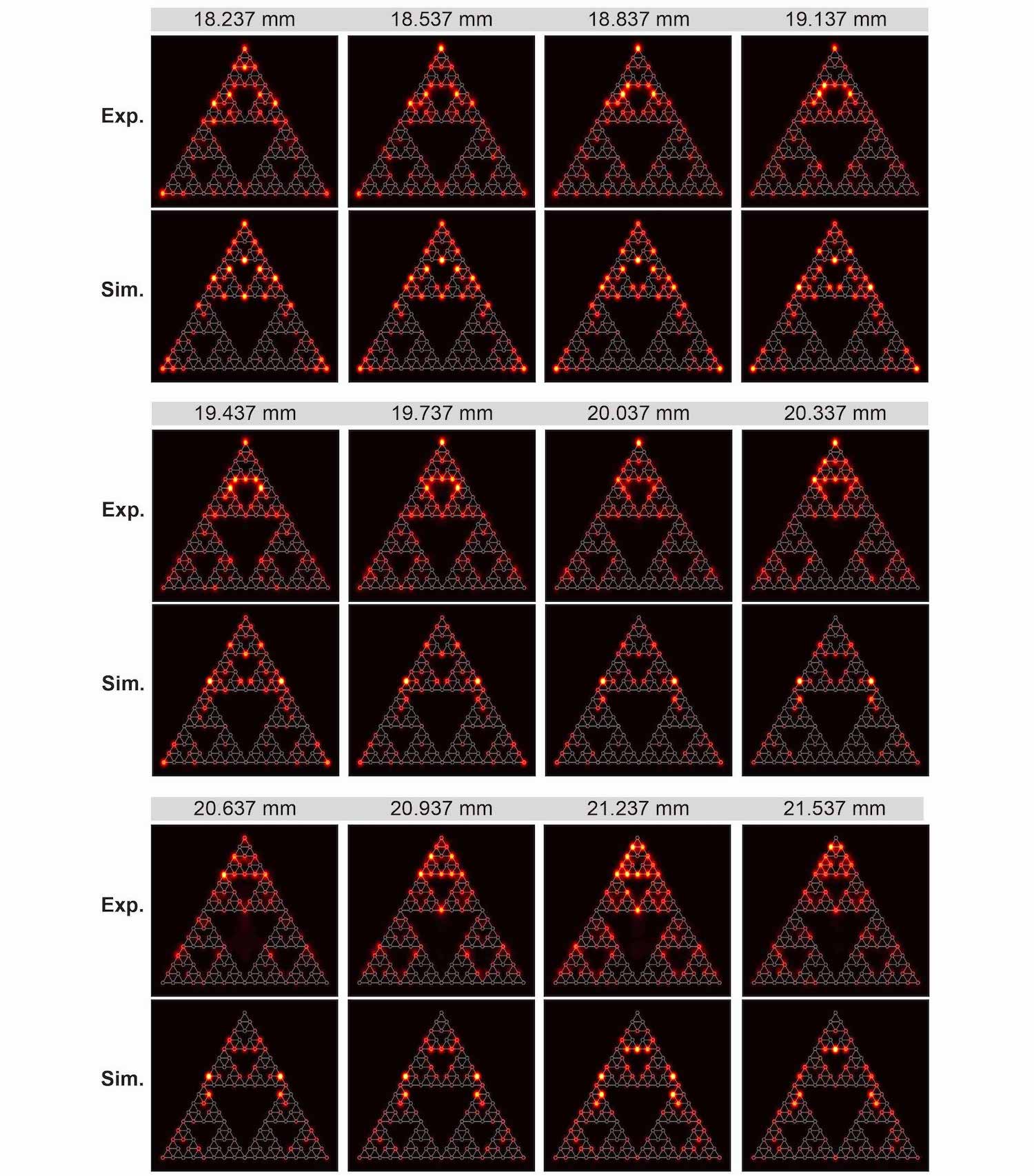}
\caption{\textbf{The evolution patterns in the Sierpi\'nski  gasket.} The propagation length starts from 18.237 mm to 21.537 mm, increasing with an increment of 0.3 mm. Experimental (Exp.) and simulated (Sim.) results are compared.}
\label{figS5}
\end{figure*}

\begin{figure*}[htp]
\centering
\includegraphics[width=0.95\columnwidth]{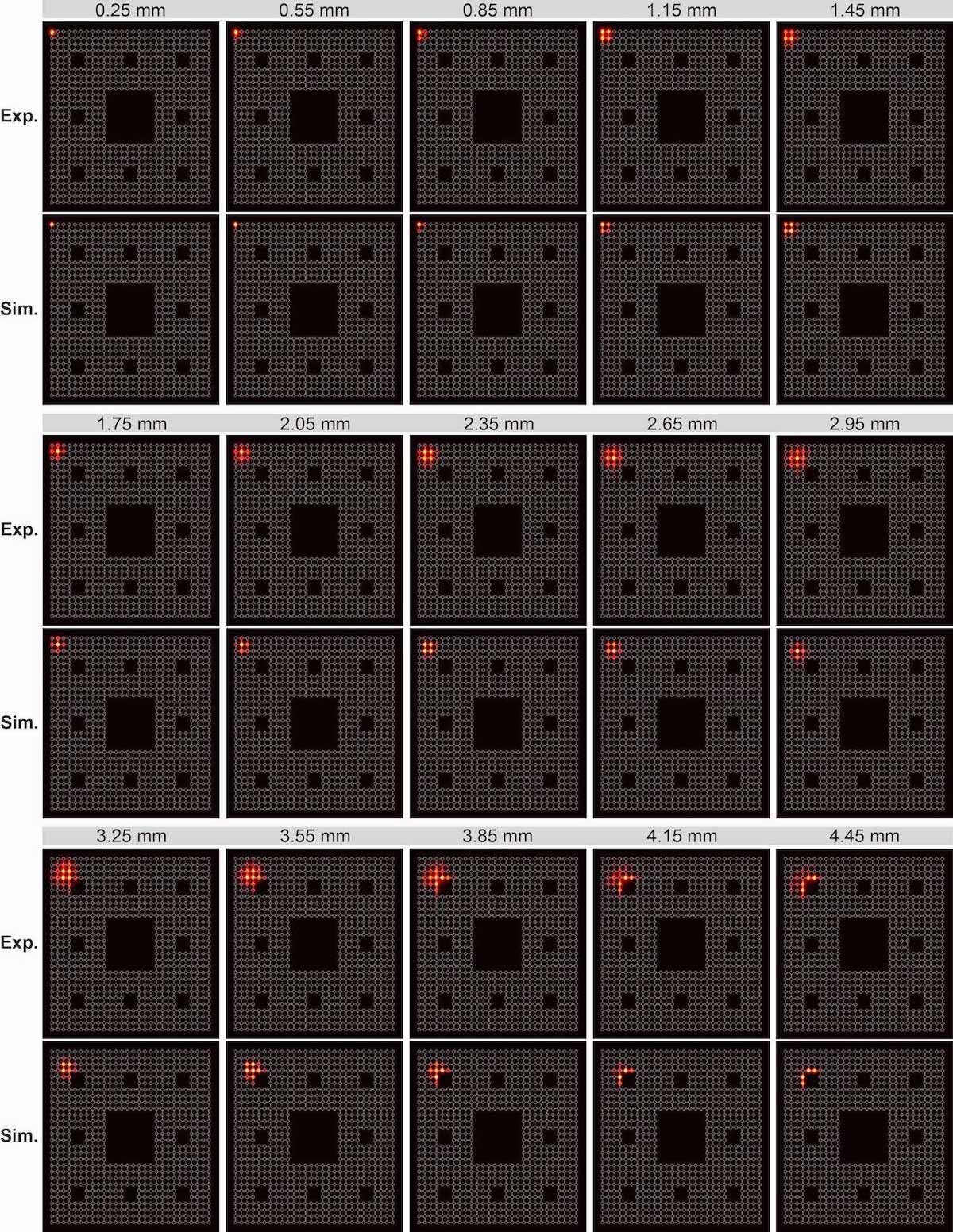}
\caption{\textbf{The evolution patterns in the Sierpi\'nski  carpet.} The propagation length starts from 0.25 mm to 4.45 mm, increasing with an increment of 0.3 mm. Experimental (Exp.) and simulated (Sim.) results are compared.}
\label{figS6}
\end{figure*}

\begin{figure*}[htp]
\centering
\includegraphics[width=0.95\columnwidth]{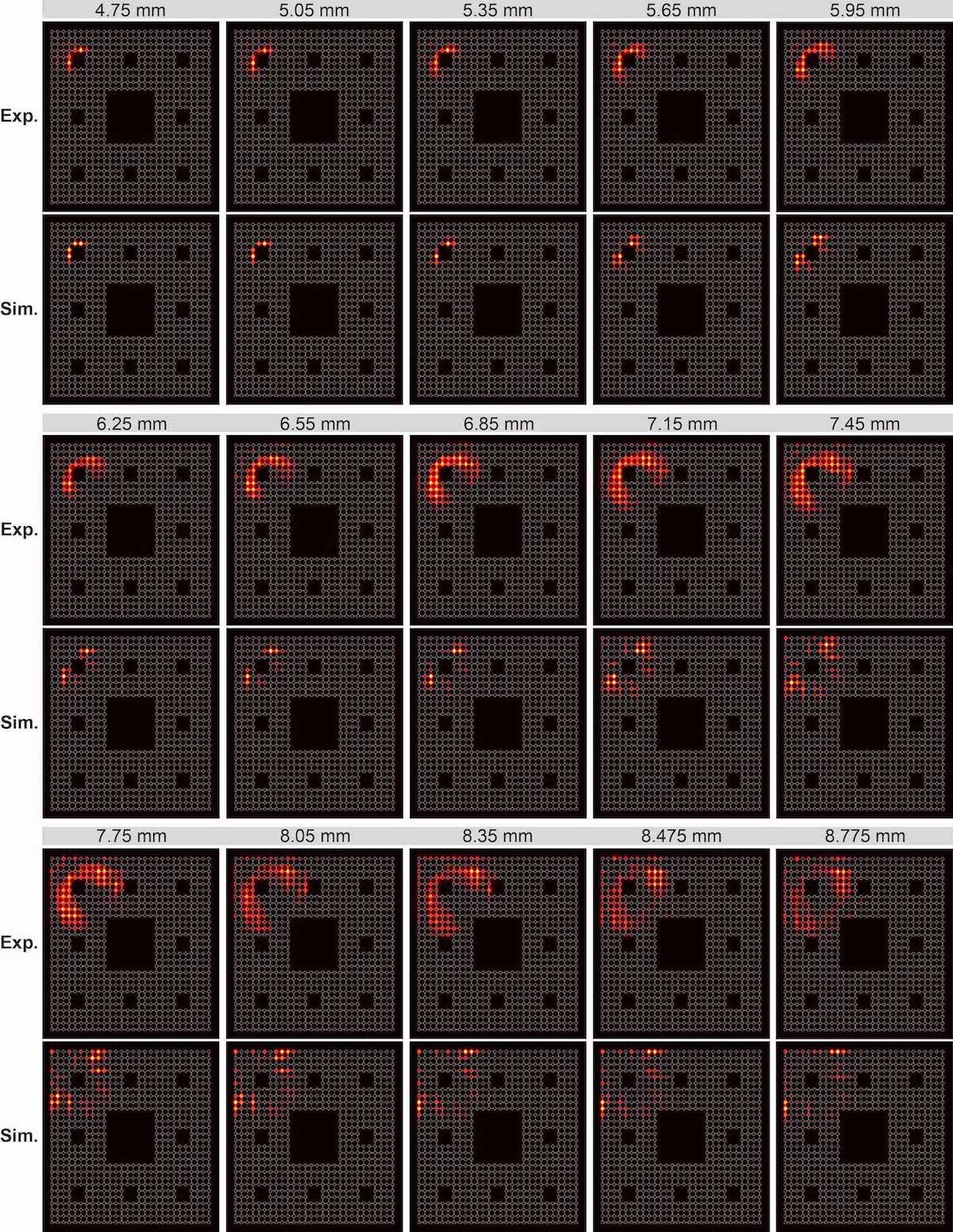}
\caption{\textbf{The evolution patterns in the Sierpi\'nski  carpet.} The propagation length starts from 4.75 mm to 8.775 mm, increasing with an increment of 0.3 mm when it is smaller than 8.35 mm and larger than 8.475 mm. The  break of regular increase at 8.35 mm is attributed to the fact that the photonic chips suffer different losses in length during the mechanical grinding processes. Experimental (Exp.) and simulated (Sim.) results are compared.}
\label{figS7}
\end{figure*}

\begin{figure*}[htp]
\centering
\includegraphics[width=0.95\columnwidth]{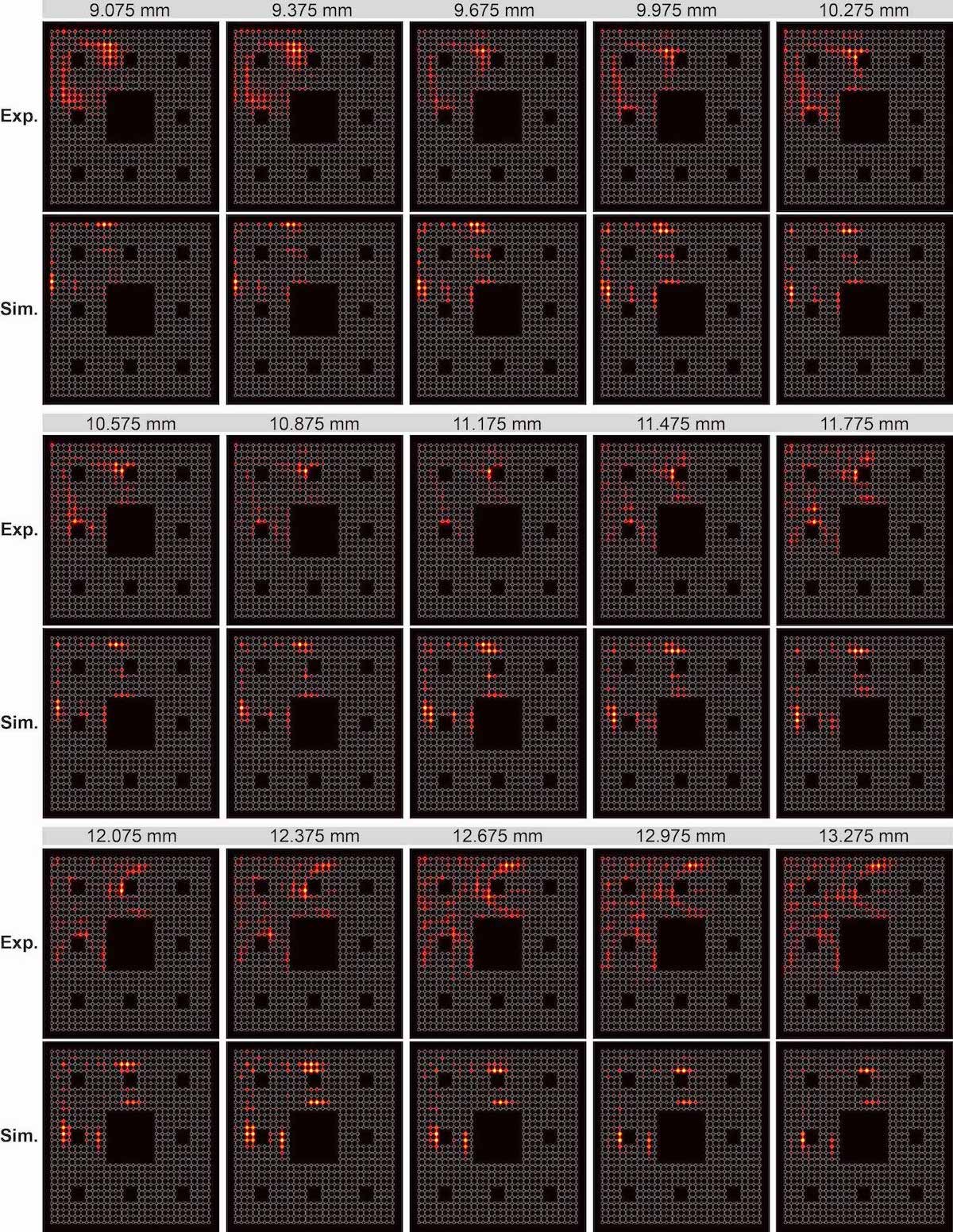}
\caption{\textbf{The evolution patterns in the Sierpi\'nski  carpet.} The propagation length starts from 9.075 mm to 13.275 mm, increasing with an increment of 0.3 mm. Experimental (Exp.) and simulated (Sim.) results are compared.}
\label{figS8}
\end{figure*}

\begin{figure*}[htp]
\centering
\includegraphics[width=0.95\columnwidth]{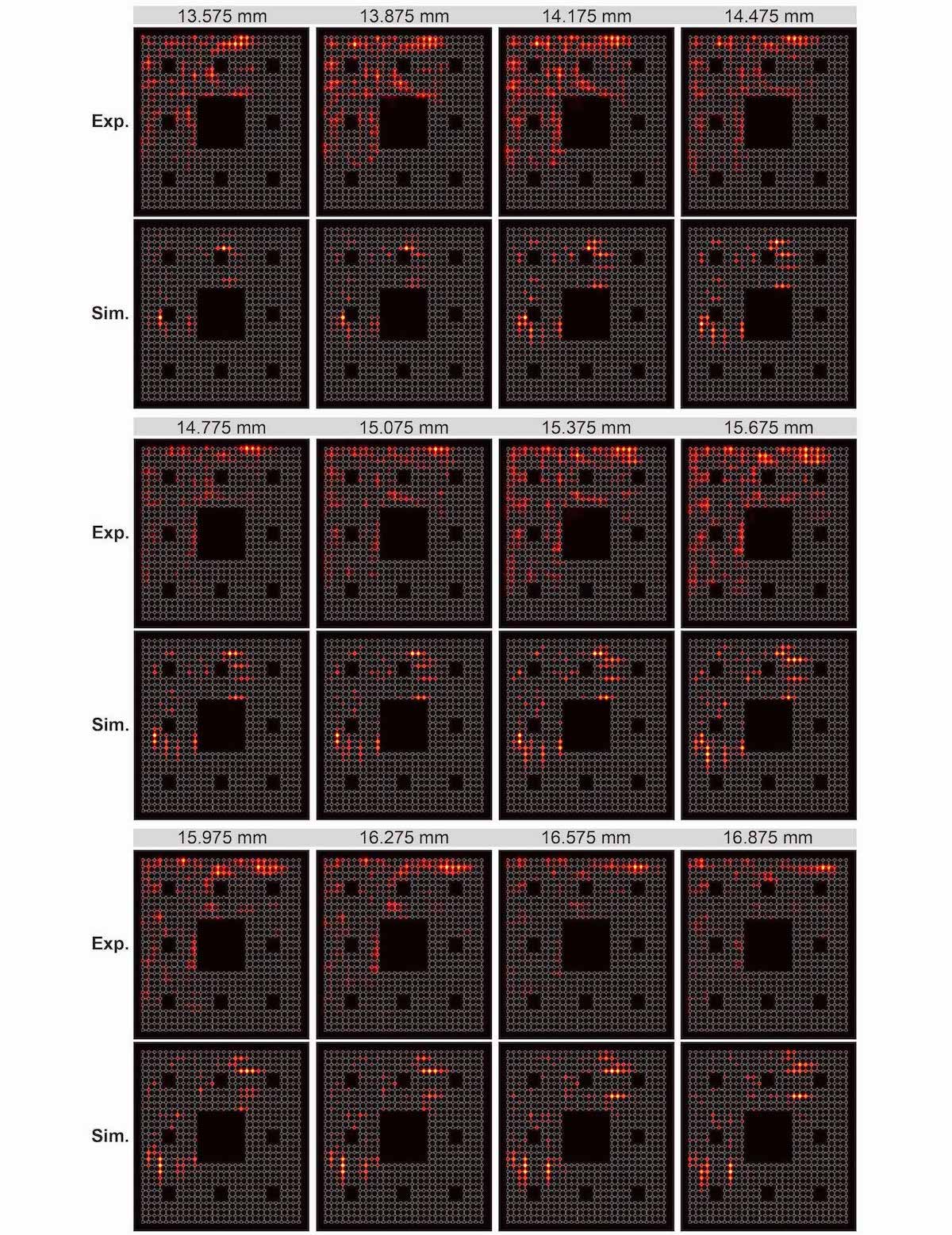}
\caption{\textbf{The evolution patterns in the Sierpi\'nski  carpet.} The propagation length starts from 13.575 mm to 16.875 mm, increasing with an increment of 0.3 mm. Experimental (Exp.) and simulated (Sim.) results are compared.}
\label{figS9}
\end{figure*}

\begin{figure*}[htp]
\centering
\includegraphics[width=1\columnwidth]{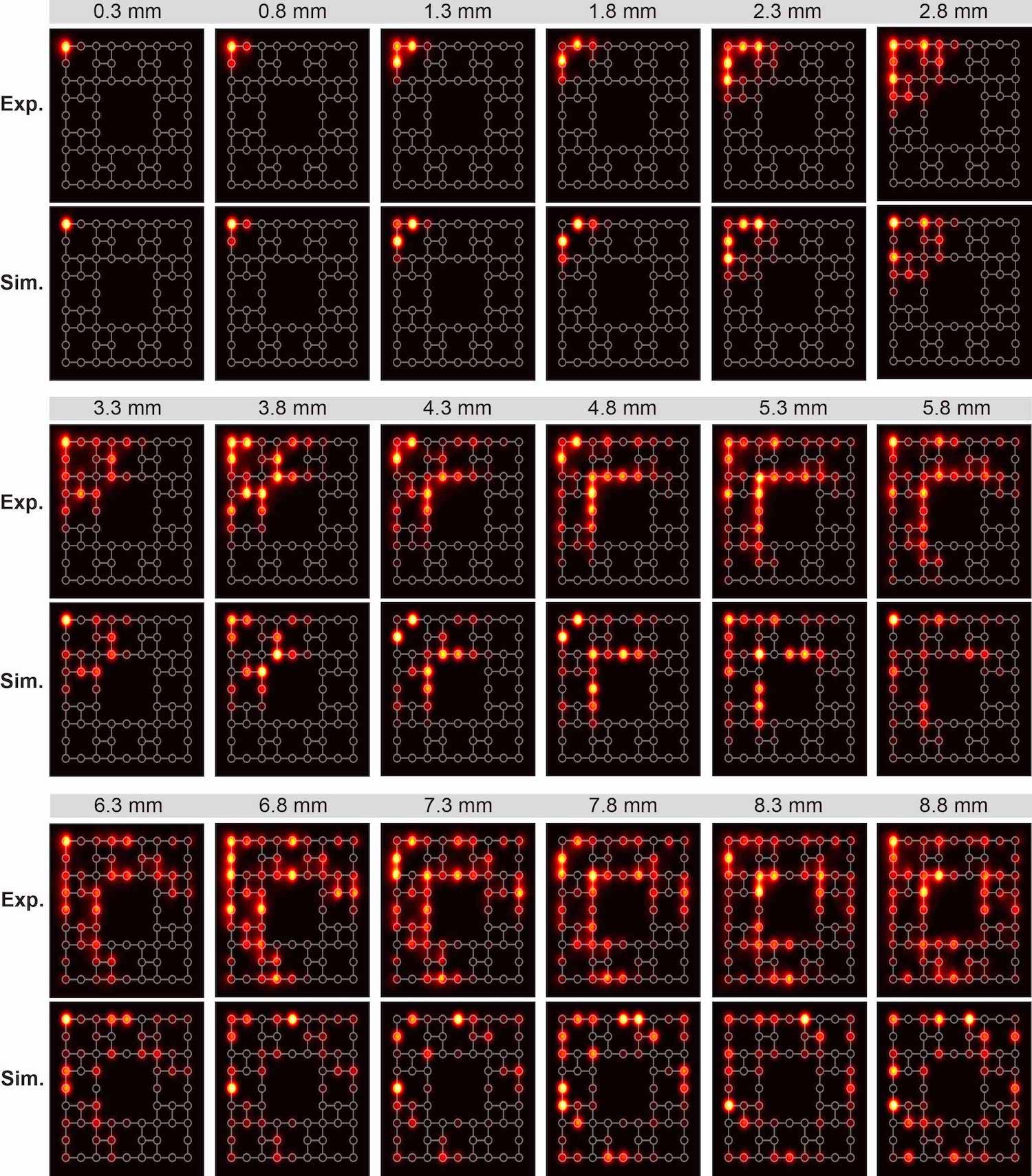}
\caption{\textbf{The evolution patterns in the dual Sierpi\'nski  carpet.} The propagation length starts from 0.3 mm to 8.8 mm, increasing with an increment of 0.5 mm. Experimental (Exp.) and simulated (Sim.) results are compared.}
\label{figS10}
\end{figure*}

\begin{figure*}[htp]
\centering
\includegraphics[width=1\columnwidth]{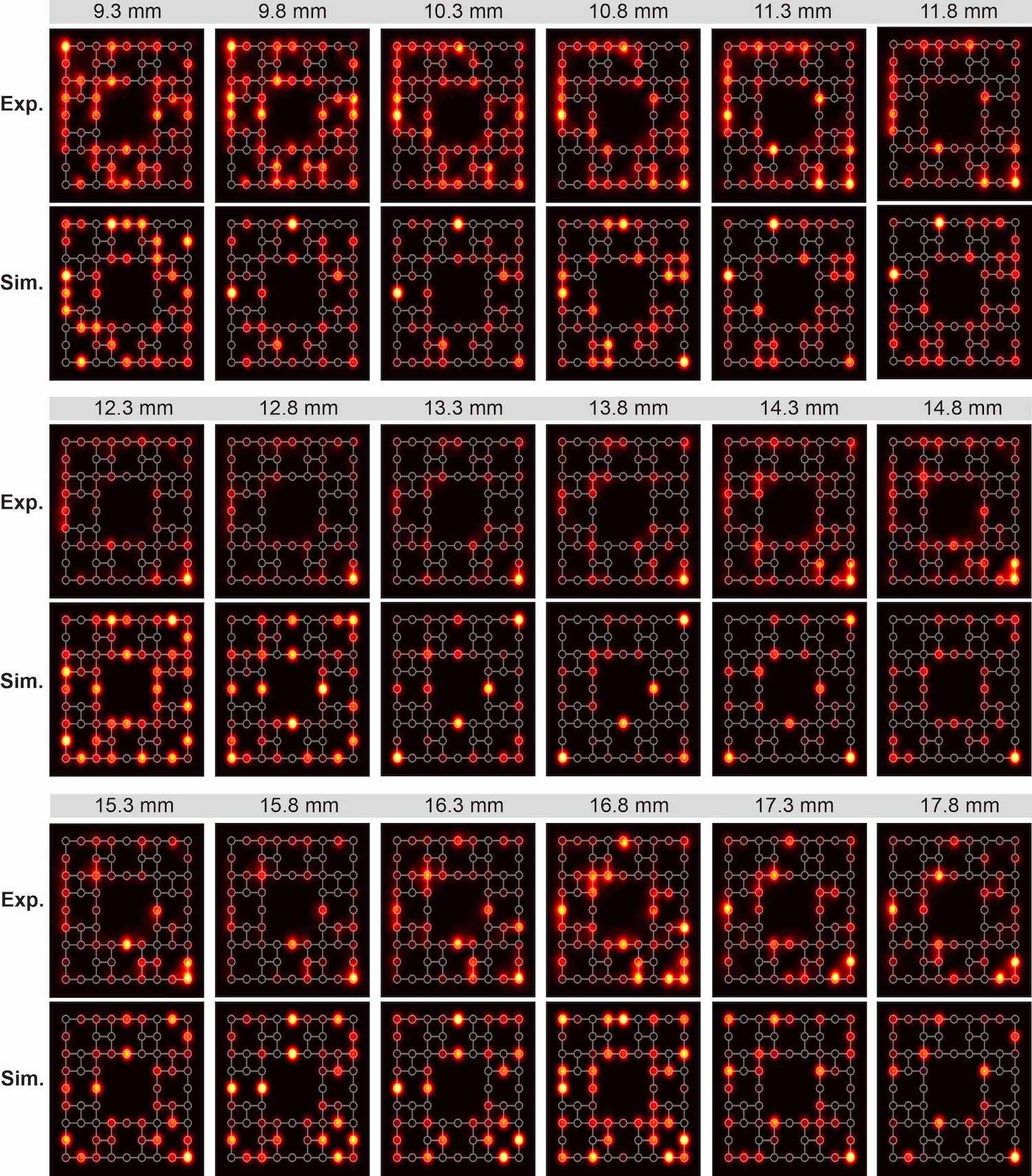}
\caption{\textbf{The evolution patterns in the dual Sierpi\'nski  carpet.} The propagation length starts from 9.3 mm to 17.8 mm, increasing with an increment of 0.5 mm. Experimental (Exp.) and simulated (Sim.) results are compared.}
\label{figS11}
\end{figure*}

\begin{figure*}[htp]
\centering
\includegraphics[width=1\columnwidth]{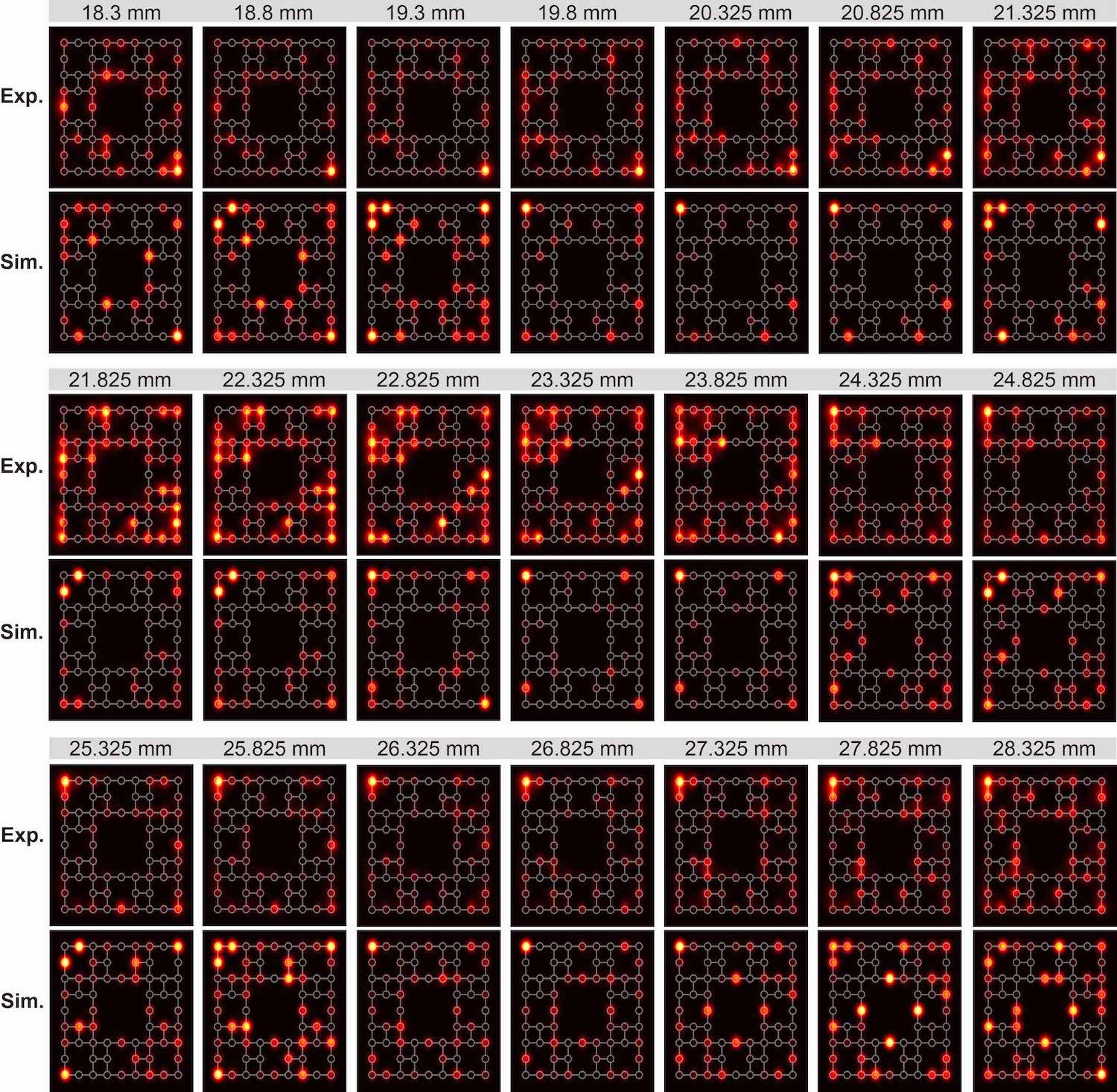}
\caption{\textbf{The evolution patterns in the dual Sierpi\'nski  carpet.} The propagation length starts from 18.3 mm to 28.325 mm and increases regularly with an increment of 0.5 mm when it is smaller than 19.8 mm and larger than 20.325 mm.  The  break of regular increase at 19.8 mm is attributed to the fact that the photonic chips suffer different losses in length during the mechanical grinding process. Experimental (Exp.) and simulated (Sim.) results are compared.}
\label{figS12}
\end{figure*}

\begin{figure*}[htp]
\centering
\includegraphics[width=1\columnwidth]{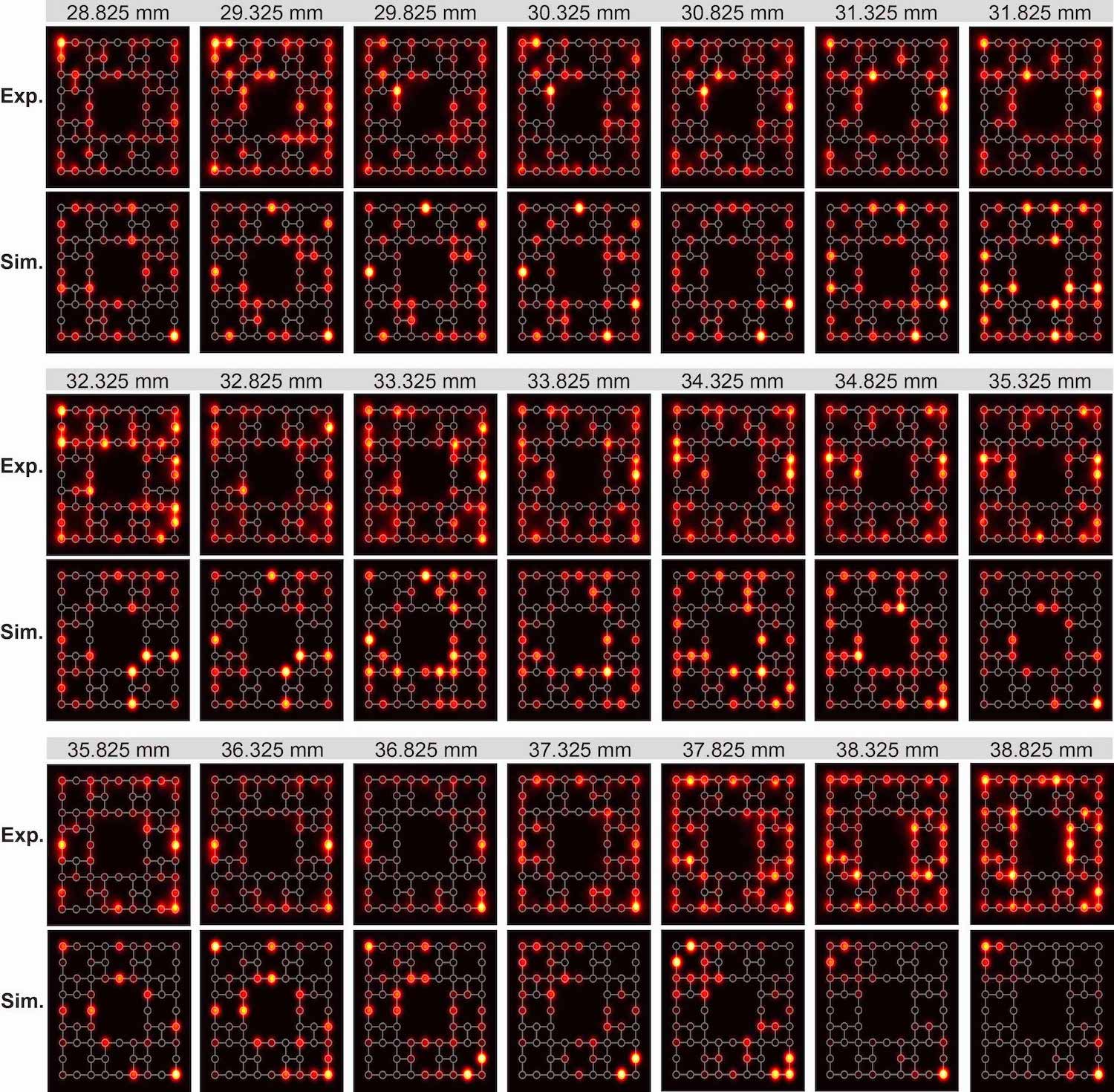}
\caption{\textbf{The evolution patterns in the dual Sierpi\'nski  carpet.} The propagation length starts from 28.825 mm to 38.825 mm, increasing with an increment of 0.5 mm. Experimental (Exp.) and simulated (Sim.) results are compared.}
\label{figS13}
\end{figure*}

Photons are injected into the photonic lattices for the Sierpi\'nski  gasket, the Sierpi\'nski  carpet and the dual Sierpi\'nski  carpet, and the evolution patterns are collected with a charge-coupled device. Photonic lattices with identical geometry but different propagation lengths are prepared, providing a platform for observing the quantum transport dynamics in the fractal networks. The propagation length  increases in constant increments, allowing a series of measurements of the evolution results at different moments. The increments are properly selected to resolve the entire quantum transport processes in different fractals, according to the simulated results. Starting from the input site, photons undergo early spreading, hindrance from the voids and reflections from the boundaries. The entire evolution patterns for the Sierpi\'nski  gasket, the Sierpi\'nski  carpet and the dual Sierpi\'nski  carpet are shown in Figs. \ref{figS1}-\ref{figS5},  Figs. \ref{figS6}-\ref{figS9} and Figs. \ref{figS10}-\ref{figS13}, respectively, unveiling the transport dynamics with unprecedented details. \\

%

\newpage
\subsection{Quantum transport in finite regular lattices}
In the cases of the Sierpi\'nski gasket and the Sierpi\'nski carpet, the quantum transport at early stage follows a normal regime and the critical point where the transport starts to deviate from the normal regime is closely related to the geometry of the photonic lattices. As mentioned in the main text, the normal regime is defined to describe quantum transport in finite regular lattices. Here we construct finite regular lattices corresponding to the studied fractals, respectively, and simulate the quantum transport in the constructed lattices, as shown in Figs. \hyperref[figS14]{S14} and \hyperref[figS15]{S15}. 

The constructed finite regular lattices provide an excellent platform to confirm and to better understand the influences on the quantum transport from the fractal geometry, owing to the facts: \textit{(i)} When photons have not reach the first voids in the fractals, the evolution environments provided by the fractal lattices are perfectly reproduced in the constructed finite regular lattices, which allows the verification whether the transport in the fractals, described by the normal regime, is identical to the transport in the actual finite regular lattices. \textit{(ii)} Once photons reach the first voids, they are exposed to environments completely different from the ones in the regular cases. By comparing the evolution results in both cases, we are able to learn about the influences from the fractal geometry.\\

\subsubsection{Finite triangular lattices}
\begin{figure*}[htp]
\centering
\includegraphics[width=1\columnwidth]{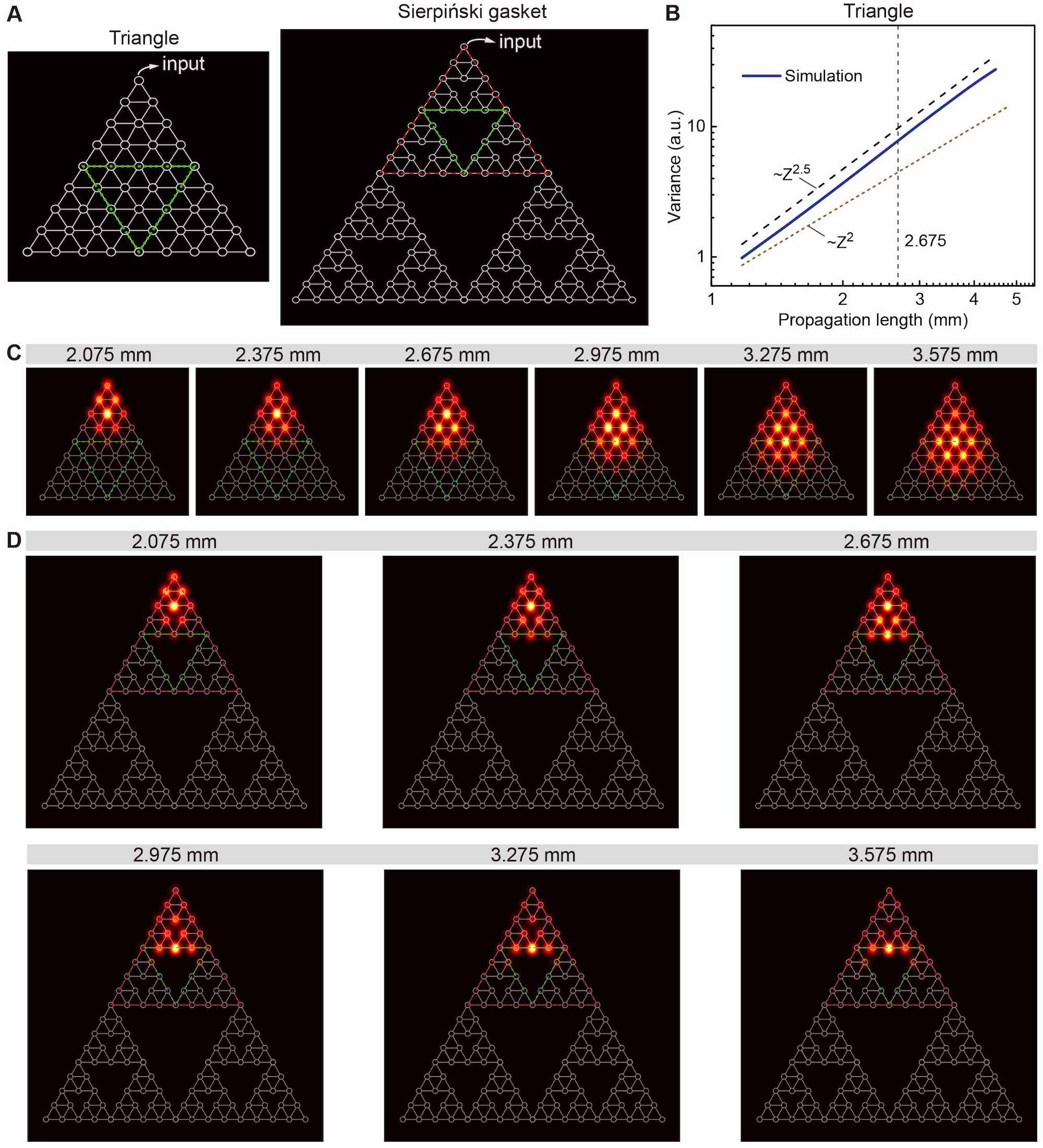}
\caption{\textbf{Quantum transport in the triangular lattice and the influence from the fractal geometry.} \textbf{(A)} The triangular lattice corresponds to the region encircled with red lines in the Sierpi\'nski gasket and is constructed by filling the void with lattice sites. The evolution of photons are simulated, with the apex site as the input.  The simulated  variance and evolution patterns for the triangle are displayed in \textbf{(B)} and \textbf{(C)}, respectively. The variance has a scaling exponent of 2.5 rather than 2, and in contrast to the Sierpi\'nski gasket, the scaling behavior holds even when the propagation length becomes larger than 2.675 mm. The evolution patterns in the triangle are compared with the ones in the Sierpi\'nski gasket, shown in \textbf{(D)}, to reveal the influence on quantum transport from the fractal geometry.}
\label{figS14}
\end{figure*}

As Fig. \hyperref[figS14]{S14A} shows, the finite triangular lattice corresponds to the region encircled by red lines in the Sierpi\'nski gasket and is constructed by filling the void with lattice sites. As defined in the main text, the white circles represent lattice sites and the white lines connecting the sites are only used for identifying the fractal geometry. The connection between two sites is solely determined by the distance between them and only the nearest sites are connected. In contrast to the diverse connectivity in the case of fractals, each internal site in the triangular lattice has a connectivity of 6, indicating its regularity. Note that the geometrical arrangement of the first five rows of the triangle is the same as the one of the Sierpi\'nski gasket, and the inputs in the two cases are both the apex sites. Therefore, before photons reach the first void or the region encircled with green lines (we call it ``the green region'' in the following), the evolution environments for the photons in the two kinds of lattices are guaranteed to be identical. We calculate the variance and depict it in Fig. \hyperref[figS14]{S14B}. It is found that when the propagation length is smaller than 2.675 mm (the point denoting the end of the normal regime in the case of the Sierpi\'nski gasket; see Fig. 2B in the main text), the scaling behavior of the variance, i.e., growing as $\sim Z^{2.5}$, for the triangle is the same as the one for the Sierpi\'nski gasket. Therefore, the quantum transport in the Sierpi\'nski gasket, described by the normal regime, is indeed identical to the transport in the triangle.  

Besides, we exhibit selected evolution patterns in the triangle at the propagation length ranging from 2.075 mm to 3.575 mm (Fig. \hyperref[figS14]{S14C}) and compare them with the corresponding patterns in the Sierpi\'nski gasket (Fig. \hyperref[figS14]{S14D}). In this interval of propagation length, the evolution in the triangle can be divided into three stages, i.e., approaching, entering and penetrating into the green region, which exactly correspond to the moments when photons encounter the first void in the Sierpi\'nski gasket. Before the propagation length increases to 2.675 mm, the main features of the patterns in the two different lattices are highly similar. However, once the propagation length becomes larger than 2.675 mm, the patterns appear to be different. It is obvious that the lack of sites in the void in the Sierpi\'nski gasket hinders the original transport of photons, leading to totally different probability distribution of photons. Moreover, with the growth of the propagation length, the influence becomes greater.\\

\newpage
\subsubsection{Finite square lattices}
\begin{figure*}[htp]
\centering
\includegraphics[width=0.9\columnwidth]{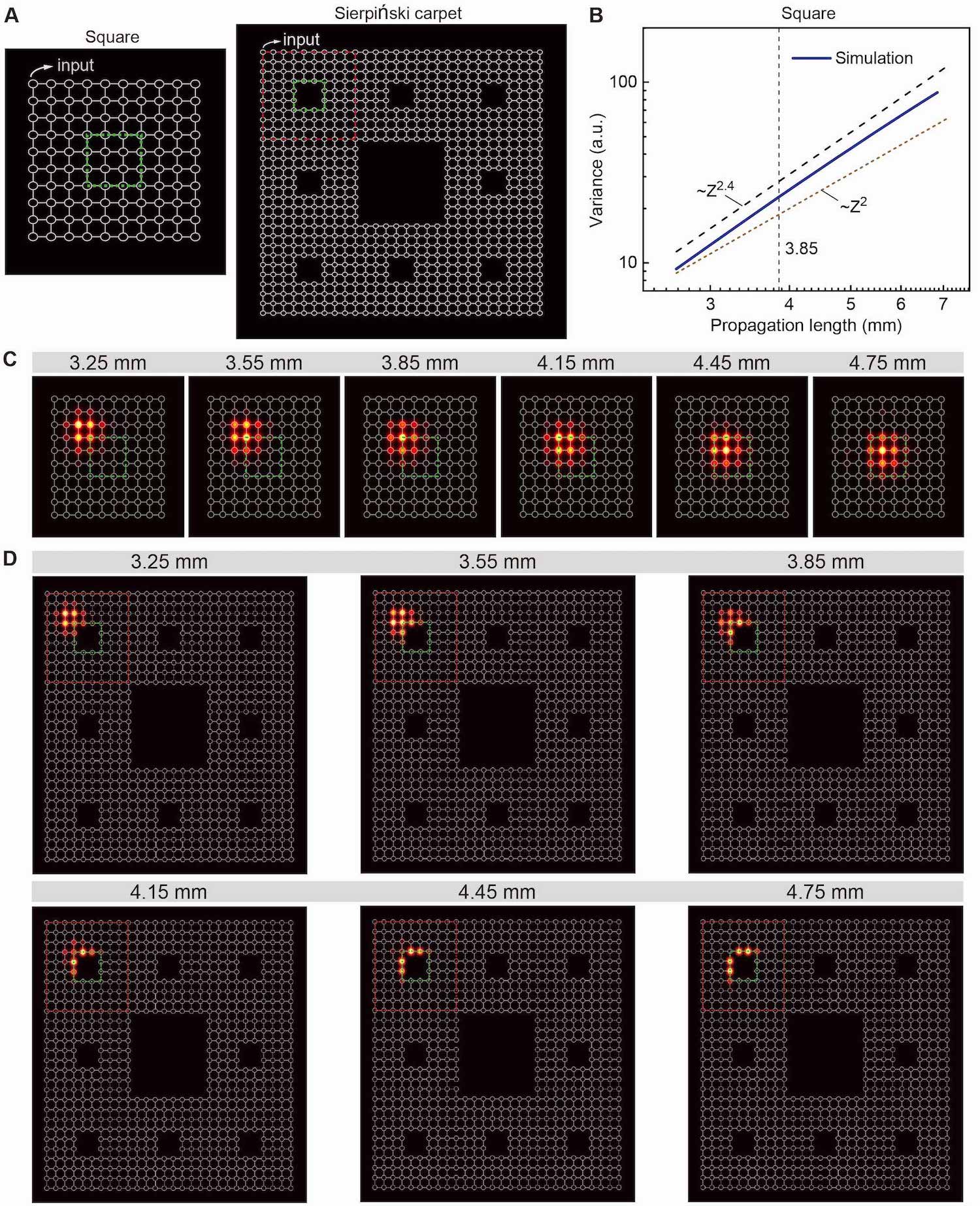}
\caption{\textbf{Quantum transport in the square lattice and the influence from the fractal geometry.} \textbf{(A)} The square lattice corresponds to the region encircled with red lines in the Sierpi\'nski carpet and is constructed by filling the void with lattice sites. The evolution of photons are simulated, with the top left site as the input. The simulated variance and evolution patterns are displayed in \textbf{(B)} and \textbf{(C)}, respectively. The variance has a scaling exponent of 2.4 rather than 2, and in contrast to the Sierpi\'nski carpet, the scaling behavior holds even when the propagation length becomes larger than 3.85 mm. The evolution patterns in the square are compared with the ones in the Sierpi\'nski carpet, shown in \textbf{(D)}, to reveal the influence on quantum transport from the fractal geometry.}
\label{figS15}
\end{figure*}

As Fig. \hyperref[figS15]{S15A} shows, the finite square lattice corresponds to the region encircled by red lines in the Sierpi\'nski carpet and is constructed by filling the void with lattice sites. Instead of diverse connectivity, each internal site in the square lattice has a connectivity of 4, indicating its regularity. Note that the geometrical arrangement of the first four rows or columns in the square are the same as the one in the Sierpi\'nski carpet and that the inputs in the two cases are both the top left corners. Therefore, before photons reach the first void or the region encircled with green lines (i.e., the green region), the evolution environments for the photons in the two kinds of lattices are guaranteed to be identical.

We calculate the variance and depict it in Fig. \hyperref[figS15]{S15B}. It is found that when the propagation length is smaller than 3.85 mm (the point indicating the end of the normal regime in the case of the Sierpi\'nski carpet; see Fig. 3B in the main text), the scaling behavior of the variance, i.e., growing as $\sim Z^{2.4}$, for the square is the same as the one for the Sierpi\'nski carpet. Therefore, the quantum transport in the Sierpi\'nski carpet, described by the normal regime, is indeed the same as the transport in the square. Besides, we exhibit selected evolution patterns in the square at the propagation length ranging from 3.25 mm to 4.75 mm (Fig. \hyperref[figS15]{S15C}) and compare them with the corresponding patterns in the Sierpi\'nski carpet (Fig. \hyperref[figS15]{S15D}). Before the propagation length increases to 3.85 mm, the main features of the patterns in the two different lattices are highly similar. However, the patterns appear to be different when the propagation length becomes larger than 3.85 mm. Obviously, the void in the Sierpi\'nski carpet hinders the original transport, leading to totally different probability distribution of photons. Moreover, with the growth of the propagation length, the influence becomes greater. \\

\newpage
\subsection{Return probability}

The return probability is the probability of a walker returning to the origin (i.e., the input site) at some moment. In the classical case, the return probability scales as $t^{-d_{s}/2},$ where $d_{s}$ is the spectral dimension of a fractal. The scaling behavior of the classical return probability has been numerically observed in the cases of Sierpi\'nski gaskets, Sierpi\'nski carpets and dual Sierpi\'nski carpets (Ref. 34). However, the law does not hold in quantum walks. In our case, the initial state is $|1\rangle$, therefore the quantum return probability can be expressed as
$$p_{1,1}(t)=|\langle 1|e^{-iHt}|1\rangle|^{2}.$$ 

\begin{figure*}[htbp]
\centering
\includegraphics[width=1\columnwidth]{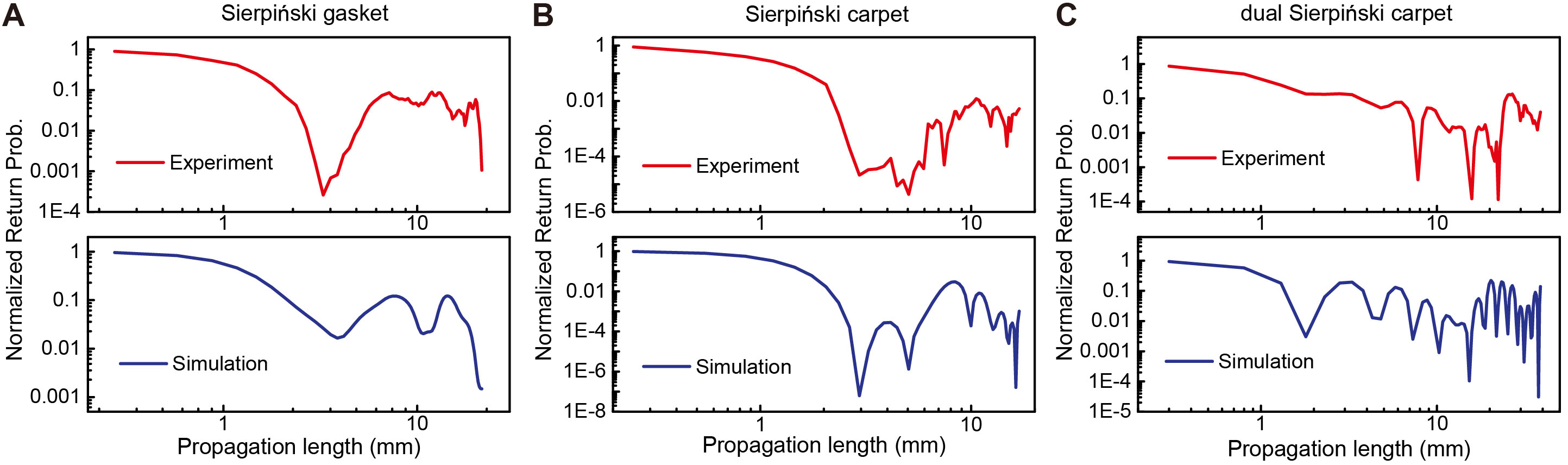}
\caption{\textbf{The quantum return probability.} \textbf{(A)} The normalized quantum return probability versus propagation length for the Sierpi\'nski gasket. The return probability first decays to a local minimum, then it shows a clear sign of oscillation. \textbf{(B)} The normalized quantum return probability versus propagation length for the Sierpi\'nski carpet. The return probability first decays to a local minimum and subsequently oscillates. \textbf{(C)} The normalized return probability versus propagation length for the dual Sierpi\'nski carpet. Though small oscillations are observed at the first stage, the return probability shows a clear sign of decaying towards a local minimum. After that, it appears to oscillate strongly. }
\label{figS16}
\end{figure*}

We depict the normalized return probability for the quantum walks in the studied fractals (Fig. \hyperref[figS16]{S16}), according to our simulated results on the probability distribution of photons. In contrast to the monotonic decay of classical return probability, which yields a line with a slope of $-d_{s}/2$ in a double-logarithmic plot, the quantum return probabilities decay to a local minimum at the first stage, then appear to oscillate strongly. Besides, the scaling behavior of the quantum return probabilities does not have a clear relation to the spectral dimension $d_{s}$. Obviously, the quantum return probabilities have completely different scaling behavior from its classical counterpart, which reveals the non-classical feature of the transport in the fractals.\\


\subsection{The P\'olya number}
\begin{figure*}[htbp]
\centering
\includegraphics[width=0.85\columnwidth]{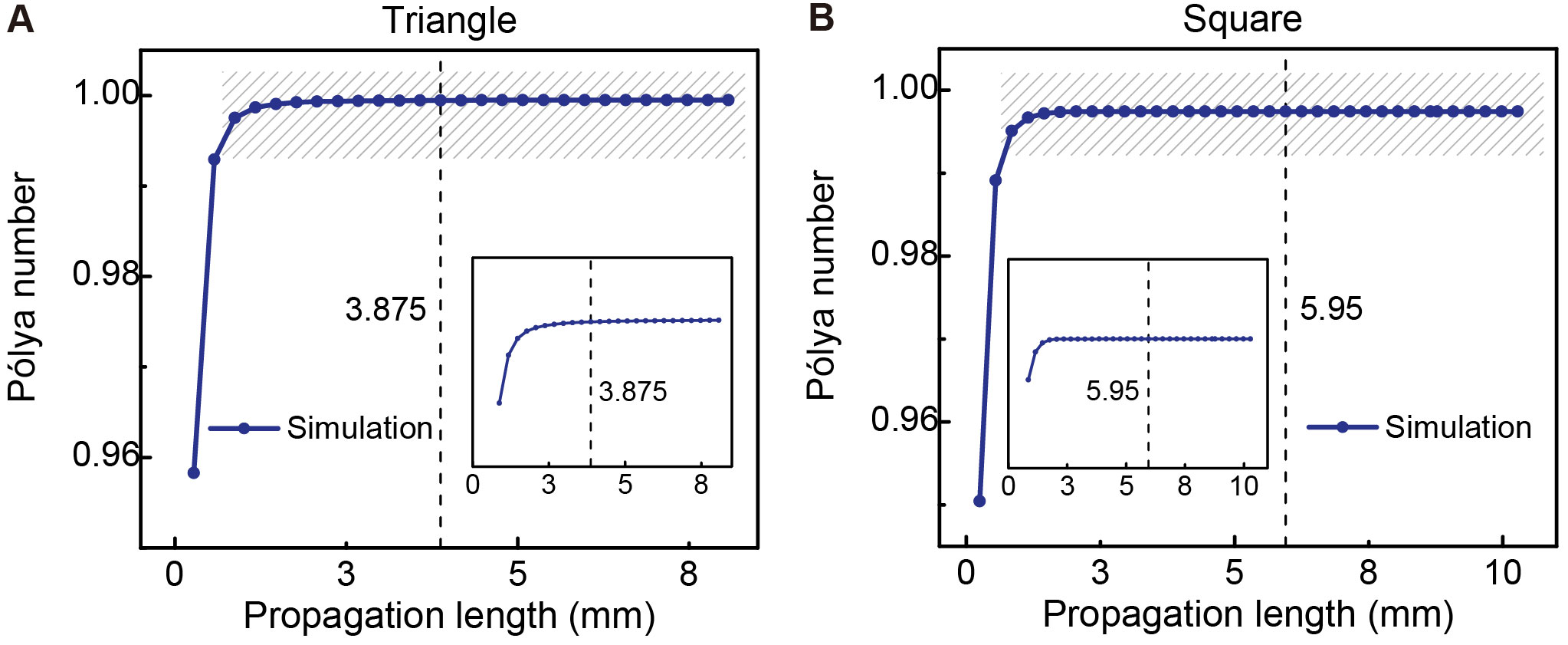}
\caption{\textbf{The P\'olya number for the regular lattices.} \textbf{(A)} The P\'olya number for the triangular lattice. It increases sharply at the early stage and then goes to a plateau. No extra growth or plateaus appear even when the propagation length becomes larger than 3.875 mm, the point indicating the beginning of the fractal regime in the Sierpi\'nski gasket. The plot highlighted with a slash pattern is magnified and displayed in the inset. \textbf{(B)} The P\'olya number for the square lattice. Similar to the case of the triangle, the P\'olya number experiences an early rapid growth and a subsequent plateau which holds even when the propagation length becomes larger than 5.95 mm, the point indicating the beginning of the fractal regime in the Sierpi\'nski carpet. The plot highlighted with a slash pattern is magnified and displayed in the inset.}
\label{figS17}
\end{figure*}

The P\'olya number, a typical characterization of a transport process, is defined as the probability of a walker ever returning to the origin during the entire transport process (i.e., the return could happen at any moment during the transport process) and it is directly related to the return probability at time $t_{i}$. If the P\'olya number is 1, it means that the walker definitely returns to the origin. Otherwise, there is a probability of the walker never going back to the origin. For continuous-time quantum walks, the P\'olya number (Ref. 53) is expressed as
$$P=1-\prod_{i=1}^{\infty}[1-p_{1,1}(t_{i})],$$
where $p_{1,1}(t_{i})$ is the return probability at time $t_{i}$ (see previous section for the definition), and the constraint $ i=1,2,..., \infty$ indicates an infinite series of time steps when we measure the evolution results. However, in general, the number of measurements on the evolution results in practical experiments is finite. Nevertheless, we plot the P\'olya number versus propagation length in Fig. \hyperref[figS17]{S17} to study its evolution during the entire transport process and to compare with the studied fractal cases. 

The P\'olya number for the finite triangular lattices (Fig. \hyperref[figS17]{S17A}) or the square lattices (Fig. \hyperref[figS17]{S17B}) shows a clear sign of saturation after a rapid growth in the beginning, which is similar to the results reported by Tang \textit{et al.} (Ref. 40). Besides, the trend of the P\'olya number for the triangle and the Sierpi\'nski gasket are highly similar before 3.875 mm, the point indicating the beginning of the fractal regime in the Sierpi\'nski gasket (see Fig. 2B in the main text). On the other hand, extra growth and plateaus appear in the P\'olya number for the Sierpi\'nski gasket after 3.875 mm, whereas in the case of the triangle these phenomena are not observed even in the magnification of the plot highlighted by a slash pattern (as shown in the inset), confirming the influence from the fractal geometry on quantum transport. As expected, we obtain the same result in the comparison between the square and the Sierpi\'nski carpet. Before 5.95 mm, the point indicating the beginning of the fractal regime in the Sierpi\'nski carpet (see Fig. 3B in the main text), the P\'olya number for the square and the Sierpi\'nski carpet have highly similar features, whereas they have obviously different trends after 5.95 mm as indicated by the plot highlighted with slash pattern and its magnification displayed in the inset.\\

\end{document}